\documentclass{aastex631} 

\usepackage{amsmath}
\usepackage{gensymb}
\usepackage{float} 
\usepackage{booktabs} 

\usepackage{newtxtext,newtxmath}
\usepackage[T1]{fontenc}
\usepackage{threeparttable}
\usepackage{booktabs,makecell,multirow}
\usepackage{longtable}
\usepackage{natbib}
\usepackage{hyperref}
\usepackage{makecell}
\usepackage{diagbox}
\DeclareRobustCommand{\VAN}[3]{#2}
\let\VANthebibliography\thebibliography
\def\thebibliography{\DeclareRobustCommand{\VAN}[3]{##3}\VANthebibliography}
\usepackage{graphicx}
\graphicspath{{./figurs/}} 

\begin{document}


\title{Single-Pulse Morphology of PSR J1935+1616 (B1933+16) Based on archival data from FAST}

\author[0009-0007-3215-2964]{R.W. Tian}
\affiliation{Guizhou Normal University, Guiyang 550001, China}

\author[0000-0002-1243-0476]{R.S. Zhao*}
\affiliation{Guizhou Normal University, Guiyang 550001, China}

\author{Marilyn Cruces}
\affiliation{Centre of Astro-Engineering, Pontificia Universidad Católica de Chile, Av. Vicuña Mackenna 4860, Santiago, Chile}
\affiliation{Joint ALMA Observatory, Alonso de Córdova 3107, Vitacura, Santiago, Chile}
\affiliation{European Southern Observatory, Alonso de Córdova 3107, Vitacura, Casilla 19001, Santiago de Chile, Chile}
\affiliation{Department of Electrical Engineering, Pontificia Universidad Católica de Chile, Av. Vicuña Mackenna 4860, Santiago, Chile}
\affiliation{Max-Planck-Institut für Radioastronomie, Auf dem Hügel 69, D-53121 Bonn, Germany}

\author{H. Liu}
\affiliation{Guizhou Normal University, Guiyang 550001, China}

\author{D. Li}
\affiliation{Department of Astronomy, Tsinghua University, Beijing 100084, China}
\affiliation{National Astronomical Observatories, Chinese Academy of Sciences, 20A Datun Road, Chaoyang District, Beĳing 100101, China}
\affiliation{Zhejiang Lab, Hangzhou, Zhejiang 311121, People’s Republic of China}

\author{P. Wang}
\affiliation{National Astronomical Observatories, Chinese Academy of Sciences, 20A Datun Road, Chaoyang District, Beĳing 100101, China}
\affiliation{Institute for Frontiers in Astronomy and Astrophysics, Beĳing Normal University, Beĳing 102206, China}

\author{C.H. Niu}
\affiliation{Central China Normal University, Wuhan 430079, China}

\author{Biping Gong}
 \affiliation{ Department of Physics, Huazhong University of Science and Technology, Wuhan 430074, China}

\author{C.C. Miao}
\affiliation{Zhejiang Lab, hangzhou 311121, China}

\author{X. Zhu}
\affiliation{Guizhou Normal University, Guiyang 550001, China}

\author[0009-0003-4524-6530]{H.W. Xu}
\affiliation{Guizhou Normal University, Guiyang 550001, China}

\author{W.L. Li}
\affiliation{Guizhou Normal University, Guiyang 550001, China}

\author{S.D. Wang}
\affiliation{Guizhou Normal University, Guiyang 550001, China}

\author{Z.F. Tu}
\affiliation{Guizhou Normal University, Guiyang 550001, China}

\author{Q.J. Zhi}
\affiliation{Guizhou Normal University, Guiyang 550001, China}

\author{S.J. Dang}
\affiliation{Guizhou Normal University, Guiyang 550001, China}

\author{L.H. Shang}
\affiliation{Guizhou Normal University, Guiyang 550001, China}

\author{S. Xiao}
\affiliation{Guizhou Normal University, Guiyang 550001, China}

\correspondingauthor{R.S. Zhao}
\email{201907007@gznu.edu.cn}


\begin{abstract}  
We utilized archived data from the Five-hundred-meter Aperture Spherical Radio Telescope (FAST) to analyze the single-pulse profile morphology of PSR J1935$+$1616 (B1933$+$16). The results show that PSR J1935$+$1616 exhibits significant micropulses as well as various changes in single-pulse profile morphology. In the FAST archived data, a total of 969 single pulses with microstructure were identified, accounting for 9.69$\%$ of the total pulse sample, with characteristic widths of $127.63^{+70.74}_{-46.25}$ $\mu$s. About half of these pulses display quasiperiodic micropulses, with a periodicity of 231.77 $\pm$ 9.90 $\mu$s. 
Among the 520 single pulses with quasiperiodic microstructure, 208 also exhibit quasiperiodicity in circular polarization, with a characteristic period of \( 244.70^{+45.66}_{-21.05} \, \mu \text{s} \). The micropulse characteristic width in circular polarization is \( 106.52 \pm 46.14 \, \mu \text{s} \).
Compared to normal pulses, the relative energy (E/<E>) of single pulse with microstructure follows a double Gaussian distribution, while that of normal pulses follows a single Gaussian distribution. 
Based on the intensity of the leading and trailing components in the single-pulse profile morphology of PSR J1935+1616, we classified the pulses into four morphological modes (A, B, C, and D). The relative energy distribution of pulses in mode A is significantly different from the others, following a double Gaussian distribution, while the relative energy distributions in modes B, C, and D follow a single Gaussian distribution. Our study also suggests a possible correlation between micropulses and single-pulse profile morphology. Single pulse with micropulses are most likely to occur in mode A, while their occurrence is least likely in mode D.
\end{abstract}
\keywords{single-pulse, micropulse, morphology}
 
\section{INTRODUCTION} \label{title_1}

Pulsars are extraordinary celestial objects characterized by their rapid rotation, high density, and strong magnetic fields \citep{lyne2012pulsar}. For some pulsars, the emission behavior shows significant variations, such as microstructure, which is the rapid fluctuations in intensity on short timescales within a single pulse.  
Since the discovery of micropulses, more than 40 pulsars with micropulse emission were introduced in different works \citep{Hankins..1971ApJ,Hankins..1972ApJ,Backer..1970Natur,Cordes..1990AJ,Lange..1998A&A,Popov..2002A&A,Kramer..2002MNRAS,Kuzmin..2003MNRAS,Crossley..2010ApJ,Mitra..2015ApJ,Wen..2021ApJ,Zhao..2023MNRAs,singh..2023}. For different pulsars, the widths of micropulses vary.
\cite{burstein1982reddenings} reported intensity fluctuations in PSR B1133$+$16 as small as 2.5 \(\mu\)s. \cite{hankins2003nanosecond} detected the smallest discernible features in Crab Nebula pulsars, with time scales on the order of nanoseconds. 
 \citet{Zhao..2023MNRAs} reported that the characteristic widths for PSRs J0034$-$0721 and J0151$-$0635 are 0.89 $\pm$ 0.44 ms and 1.55 $\pm$ 1.02 ms, respectively. Besides, the micropulses in some pulsars shows quasi-periodicity, which was first discovered in PSR B0950+08 by \citet{Hankins..1971ApJ}. \cite{1981IAUS...95..199B} found that in different pulsars, the quasi-periodicity has the same period across different radio frequencies. \citet{Zhao..2023MNRAs} detected quasi-periodic micropulses in 43 single pulses of PSR J0034-0721, with a typical characteristic period of approximately 2.0 ms.
 Furthermore, \cite{Popov..2002} also noted the presence of microstructures with a width of 150 \(\mu\)s in PSR B1933$+$16. \cite{Mitra..2016MNRAS} reported that the quasi-periodicity of micropulses in PSR B1933$+$16 at 1.5 GHz and 4.5 GHz was 0.4 $\pm$ 0.2 ms and 0.35 $\pm$ 0.16 ms, respectively.
 While there are so many observations, the theoretical understanding of micropulse radiation mechanisms remains incomplete (e.g., Hankins \citeyear{Hankins..1971ApJ}; Wen \citeyear{Wen..2021ApJ}).
 \cite{Petrova..2004A&A} suggested that the typical width of micropulses can be explained by relativistic plasma beams. 
\cite{1979ApJ...233..317H} studied the propagation of time-dependent polarized radiation through a cold, field-free plasma with relativistic velocity shear. They proposed that the quasi-periodic modulation of micropulses could arise from propagation effects as the radiation traverses shearing regions within the pulsar magnetosphere. 
 \cite{chain..1983.} hypothesized that strong electromagnetic waves generate nonlinear instabilities in plasma, leading to micropulses. 
\cite{Petrova..2004A&A} showed that micropulses may be caused by variable gain effects in excited Compton scattering. Thus, research on micropulse phenomena still requires a combination of observational data and theoretical analysis to provide a more comprehensive explanation and constraints. 

The morphology of single pulses is known to vary, and mode changing refers to the phenomenon in which a pulsar's mean pulse profile abruptly transitions between two or more quasi-stable emission states, with each mode lasting for different durations.
Early observations of PSRs B1237$+$25 and B0329$+$54 revealed that the relative intensity of the components of their average pulse profiles changes in different modes \citep{Backer..1970Natur,lyne1971mode,Bartel..1982ApJ,morris1981mode}. 
Another situation is that in the early study by \citep{1968Natur.220..231D}, it was observed that the position of subpulses in pulsars underwent systematic drift over time. The duration of the consistent single-pulse morphology varies \citep{rejep2022nulling}.
\cite{fowler1982pulse} reported that PSR B1822$-$09 exhibits two types of single-pulse morphologies: the quiet mode (Q-mode) and the burst mode (B-mode). In the Q-mode, the precursor is weak, drifting subpulses are strong in the main pulse, and the interpulse is relatively stronger. In the B-mode, the precursor is much stronger, drifting subpulses disappear, and the interpulse weakens.
\cite{wang2007pulsar} discovered short timescale single-pulse morphology variations in six pulsars (PSR J1843$-$0211, PSR J1703$-$4851, PSR J1701$-$3726, PSR J1648$-$4458, PSR J1326$-$6700, PSR J0846$-$3533), with durations ranging from a few seconds to a few minutes. \cite{2022ApJ...933..210M} found that the shortest duration of the same single-pulse morphology in PSR J0026$-$1955 is within 10 pulse periods. \cite{basu2023mode} found that the central bright mode in PSR B0844$-$35 occupies 10$\%$ of the observation time and typically lasts about 10 pulse periods. Currently, multiple sources exhibit characteristics of single-pulse morphology changes. The integrated profiles of the same single-pulse morphology over longer timescales in some pulsars are useful for constructing the geometric structure of the polar cap and for providing additional constraints on pulsar radiation mechanisms (e.g., Rankin \citeyear{Rankin..1983ApJ}; Lyne $\&$ Manchester \citeyear{1988MNRAS.234..477L}). 
\cite{Bartel..1982ApJ} suggested that changes in single-pulse morphology may be due to variations in the conditions of the magnetic cap region above the pulsar's surface. Meanwhile, \cite{wang2007pulsar} proposed that the redistribution of currents in the pulsar magnetosphere leads to changes in the emission mode of the radio beam, thereby causing changes in the pulse profile. \cite{wen2016investigation} suggested that different subpulse drift modes indicate possible changes in emission properties and may suggest local changes in the electromagnetic field configuration in the gap.

In this paper, we investigate the microstructure of single pulses and the morphological variations of individual pulses in PSR J1935+1616 based on high-sensitivity FAST observations. In Section 2 we describe the FAST data of this pulsar and the data processing methods we used. In Section 3, we analyzed the microstructure and the changes in single-pulse morphology modes of PSR J1935$+$1616. In Section 4, we further discuss our findings and the properties associated with them. Finally, in Section 5, we summarise the whole research process and draw final conclusions.

\clearpage
\section{DATA PROCESSING} \label{title_2}

PSR J1935$+$1616 was observed on September 19, 2019, using FAST in tracking mode with the central beam of its 19-beam receiver. The receiver operated at a center frequency of 1250~MHz with a bandwidth of 500~MHz. Leveraging its innovative design, which incorporates 1,000 actuators for precise pointing and tracking, as well as the high sensitivity of its 19-beam L-band receiver, FAST enables simultaneous data acquisition for pulsar studies, neutral hydrogen detection, and transient searches \citep{2011IJMPD..20..989N,2018IMMag..19..112L}. The observations were recorded in 8-bit PSRFITS search mode format, with 4096 frequency channels and a sampling time of 49.512~$\mu$s. A total of 9998 single pulses were detected during the one-hour observation. Single-pulse stacks were subsequently produced from the de-dispersed data using \textsc{DSPSR} \citep{van.Straten..2011PASA} and the ephemeris of PSR J1935$+$1616 obtained from \textsc{PSRCAT} \citep{Manchester..2005AJ}. To remove radio frequency interference (RFI), the \textsc{paz} tool in the \textsc{PSRCHIVE} package \citep{Hotan.2004PASA} was applied to the single pulses in the frequency domain.

\begin{figure}[ht!]
 \centering
\includegraphics[width=0.8\textwidth,height=0.45\textheight]{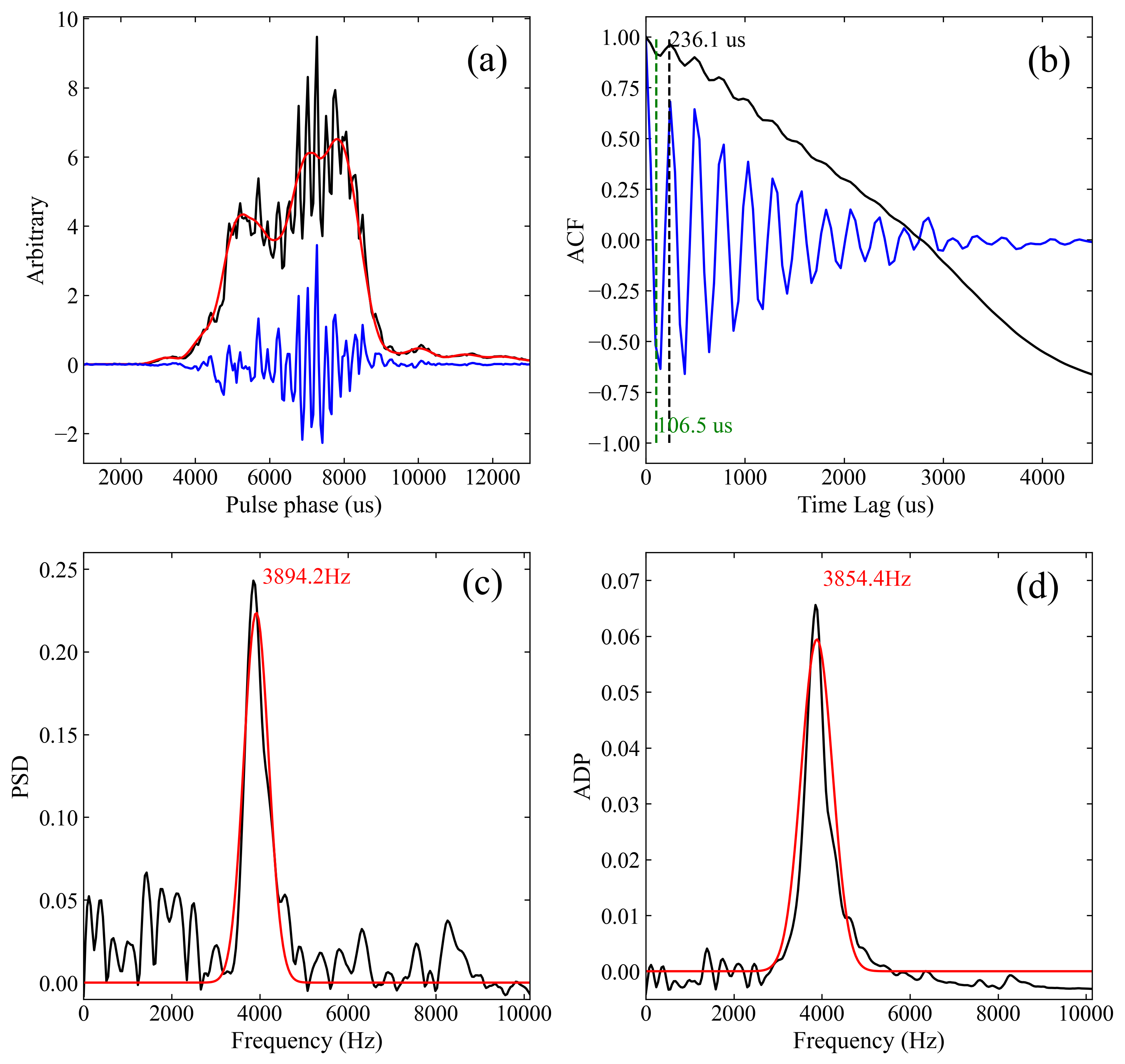}
\caption{Pulse No.789:
    (a) The black solid line shows the profile of the single pulse, the red solid line represents the smoothed envelope of the corresponding subpulse, and the blue solid line represents the residual after subtracting the smoothed envelope from the single pulse profile.
    (b) The relationship between the autocorrelation function (ACF) of the single pulse and the residuals with the time lag is shown. The black solid line represents the ACF of the single pulse, and the blue solid line represents the ACF of the residuals. The green dashed line and the black dashed line mark the troughs and peaks of the ACF, respectively. These markers help identify periodic features.
    (c) The black solid line represents the power spectrum of the residuals, and the red solid line represents its Gaussian fit. The peak of the power spectrum is at 3894.2 Hz, indicating significant periodic components at this frequency.
    (d) The black solid line represents the power spectrum of the residuals' ACF, and the red solid line represents its Gaussian fit. The peak of this power spectrum is at 3854.4 Hz, further confirming the periodic components in the residuals.}
\label{pulse:789}
\end{figure}

\begin{figure*}
    \begin{minipage}{0.5\linewidth}
        \centering
        \includegraphics[width=0.98\textwidth]{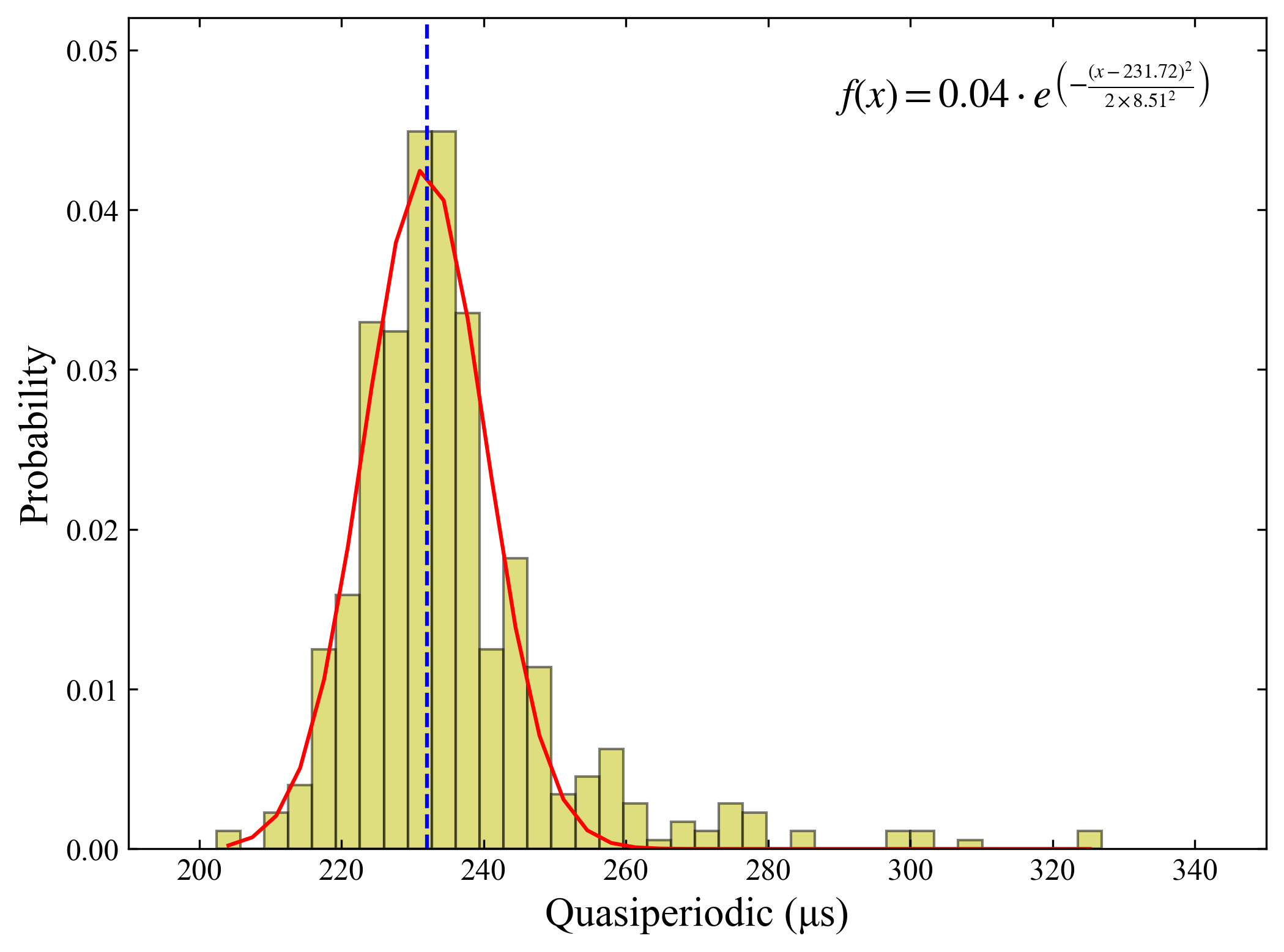} 
        \par (a) 
    \end{minipage}%
    \hspace{\fill} 
    \begin{minipage}{0.5\linewidth}
        \centering
        \includegraphics[width=0.98\textwidth]{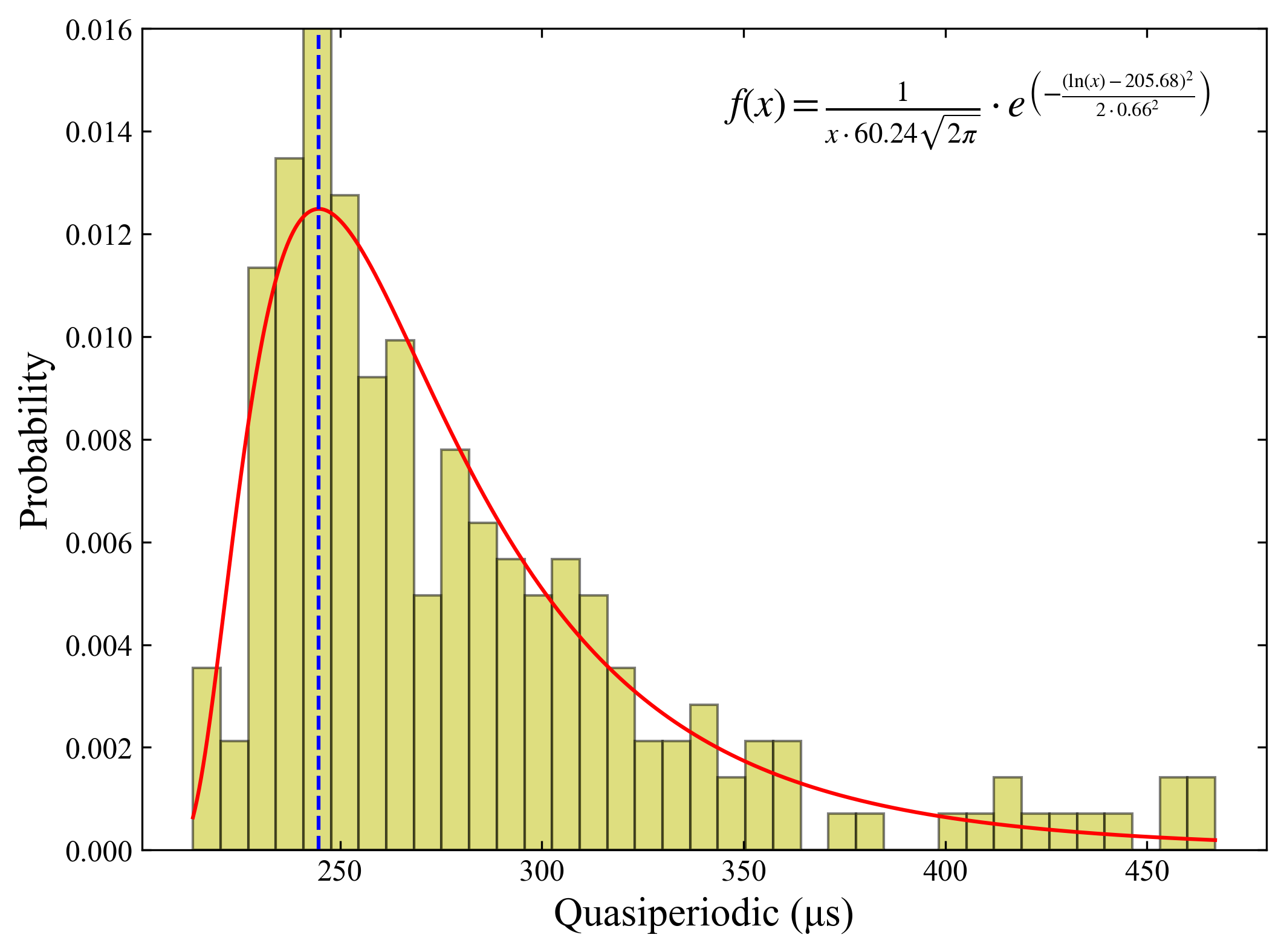} 
        \par (b) 
    \end{minipage}
    \caption{ (a) The histograms illustrate the periodicity distribution of microstructures based on 520 single pulses containing quasiperiodic micropulses. 
    The red solid line represents the best Gaussian fit. (b) The red solid line represents the best logarithmic function fit, showing the quasiperiodic distribution of 207 QMPs in circular polarization. In both panels, the blue dashed line marks the peak position of the fitted curve, with the corresponding fit formula and parameters displayed in the top right corner.
    }
    \label{Quasiperiodic}
\end{figure*}

\begin{figure*}
    \begin{minipage}{0.5\linewidth}
        \centering
        \includegraphics[width=\textwidth]{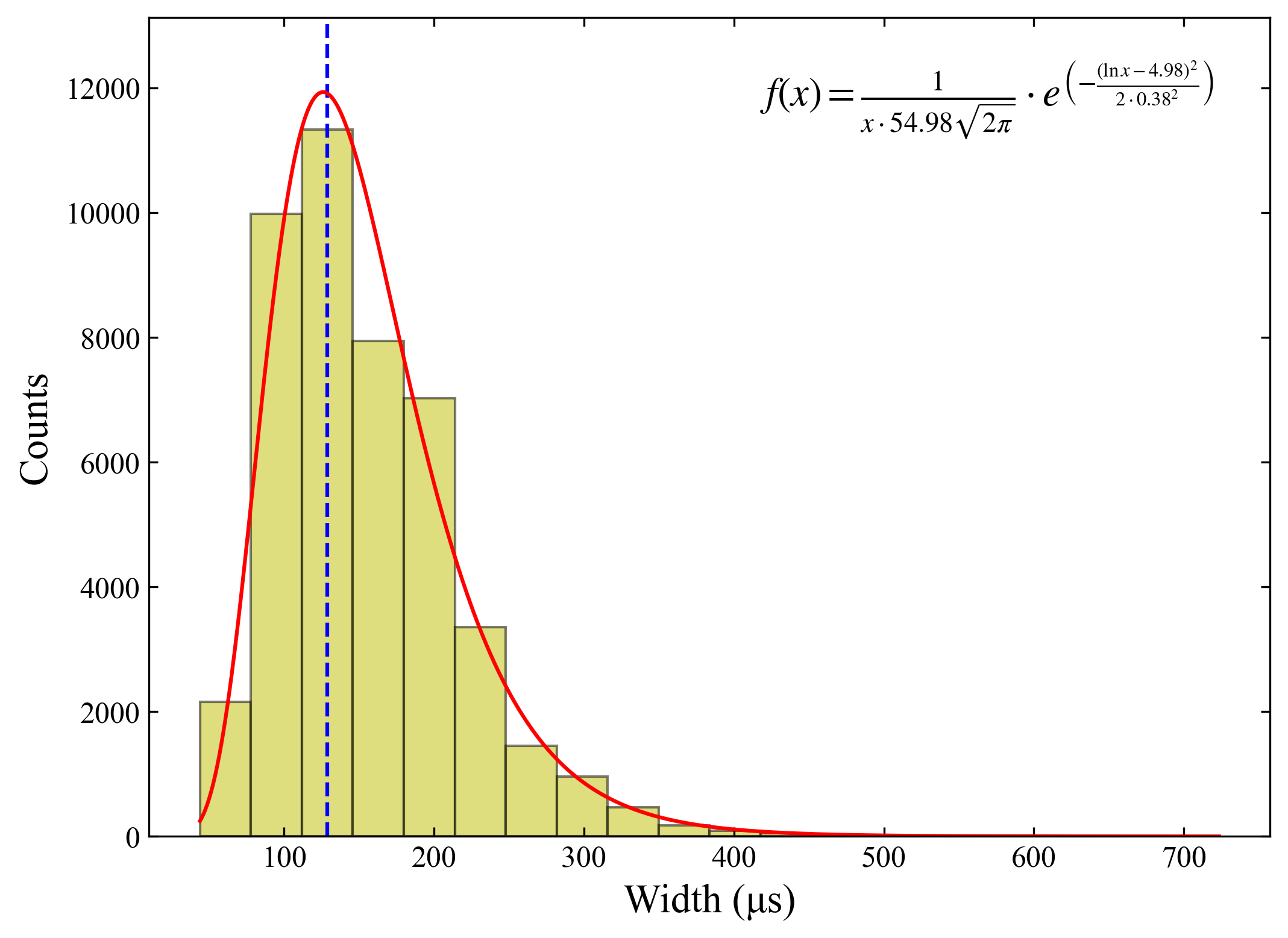}
        \centerline{(a)}
    \end{minipage}%
    \begin{minipage}{0.5\linewidth}
        \centering
        \includegraphics[width=\textwidth]{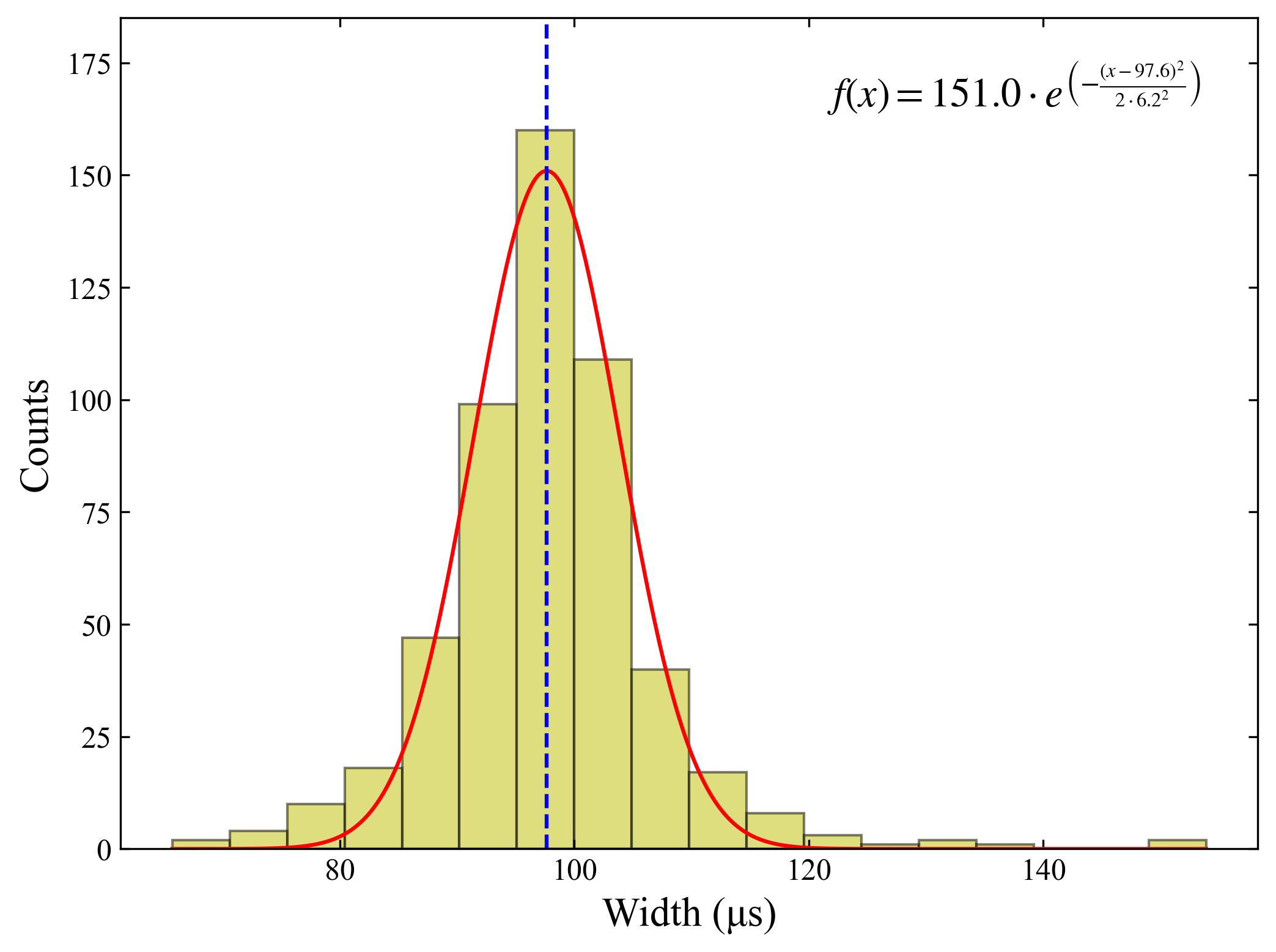}
        \centerline{(b)}
    \end{minipage}
    \caption{
    (a) The width distribution of micropulses in PSR J1935$+$1616, with the red solid line representing the log-normal distribution fit and the blue dashed line indicating the peak position. (b) The width distribution of micropulses for 520 single pulses with quasi-periodic micropulses. The red solid line represents the Gaussian fit, and the blue dashed line indicates the peak position. For both panels, the top right corner displays best log-normal (a) and Gaussian (b) fit, respectively.}
  \label{width}
\end{figure*}

\clearpage
\section{ANALYSIS \& RESULTS} \label{title_3}
\subsection{Analysis of Microstructure, Polarization, and Energy Distribution in J1935+1616} 
\label{title_3.1}
\subsubsection{ Microstructure width measurement, microstructure quasi-periodicity in total intensity, and circular polarization}  \label{title_3.1.1}

Fllowing the method in \citet{Zhao..2023MNRAs}, We have identified 969 single pulses exhibiting microstructure and 520 single pulses exhibiting quasi-periodic microstructure, accounting for approximately 9.69$\%$ and 5.20$\%$ of all pulses, respectively. For convenience in the following analysis, MP is used to represent single pulses with microstructure, and QMP represents single pulses with quasi-periodic microstructure. To determine the emission timescales of microstructures, we processed the data using autocorrelation functions (ACF) and fast Fourier transforms. Taking the No.789 pulse as an example (see Figure \ref{pulse:789}), in panel (a), each individual pulse ( black curve) was smoothed to the red curve with a fifth-order polynomial fitting, and the blue line gives the residuals of them. In panel (b), the black and blue curves represent the ACF of the individual pulse and the residuals, respectively. The ACF curve of the residuals shows a change in slope at 106.5 \(\mu\)s, indicating the presence of microstructures. Additionally, it exhibits synchronous and quasi-periodic oscillations with intervals of approximately 236.1 \(\mu\)s. To obtain reliable quasi-periods, the power spectrum of the data (PSD) and the power spectrum of the ACF derivative (ADP) was performed (shown as the black curves as shown in panels (c) and (d)) \citep{Lange..1998A&A}. The red curves represent their Gaussian fits, with PSD and ADP showing prominent peaks at 3894.2 Hz and 3854.4 Hz, respectively, which are consistent with the quasi-periodic values estimated by the ACF. 
We conducted a statistical analysis of the periods with quasi-periodic microstructures (see panel a in Figure \ref{Quasiperiodic}). The results show that these pulses follow a single Gaussian distribution with a characteristic period of 231.77 $\pm$ 9.90 \(\mu\)s. Compared to the quasi-period of 400 $\pm$ 200 \(\mu\)s reported by \cite{Mitra..2016MNRAS}, our result is significantly shorter. Furthermore, among the 520 single pulses exhibiting quasi-periodic micropulses in intensity, 208 showed characteristics of circularly polarized quasi-periodic micropulses.
Among the 520 single pulses with quasiperiodic micropulses, 208 exhibit characteristics of quasiperiodic micropulses in circular polarization. Compared to total intensity, the circular polarization intensity is typically much weaker. The weaker circular polarization signal may lead to uncertainties in the measurement of its timescale and periodicity. Additionally, the inconsistency between the timescale of the microstructure and the timescale of the total intensity microstructure could also contribute to this result.
We performed a statistical analysis of these circularly polarized quasi-periodic micropulses (see panel b in Figure \ref{Quasiperiodic}), and obtained a confidence interval for their characteristic period of $244.70^{+45.66}_{-21.05}$ $\mu$s. This is consistent with the characteristic period observed in the intensity distribution within the uncertainty range. However, due to differences between the quasi-periodic distributions of intensity and circular polarization, we applied the K-S test to assess the maximum difference (Dn) between the cumulative distribution functions (CDFs) of the two datasets, along with the corresponding significance level (P). The results showed that Dn = 0.59 > 0.05 and P = 0 < 0.05, indicating a significant difference in the distributions of the two datasets.

Figure \ref{width} shows the distribution of micropulse widths for PSR J1935$+$1616. The micropulse width was estimated using the second central moment, as described in \cite{Zhao..2023MNRAs} (see Equations \ref{eq:1} and \ref{eq:2}). The results indicate that the width distribution follows a log-normal distribution (see Figure \ref{width} (a)), with a characteristic micropulse width of $127.63^{+70.74}_{-46.25}$ $\mu$s. Additionally, 
Using the first minimum minimal of the ACF function shown in Figure
\ref{pulse:789} (b), we statistic estimate the widths of QMPs and give a characteristic width of $97.63 \pm 10.42$ $\mu$s (see Figure \ref{width} (b)).
And, based on the second central moment estimation, we calculated the micropulse characteristic width in circular polarization for 969 single pulses containing micropulses. The obtained characteristic value is \(106.52 \pm 46.14\) \(\mu\)s, with the statistical distribution following a Gaussian distribution (see Figure \ref{V_Width}).
The characteristic widths obtained from both methods are consistent within the uncertainty uncertainties. 
Since the method used in panel (b) measures the average width of all micropulses in a specific single pulse, while panel (a) plots the width of individual micropulses occurring in a single pulse, there is a difference in the distribution forms between the two methods. We further performed a K-S test to compare these distributions.
The results yielded a statistic of $D_n = 0.71 > 0.05$ and $P = 0 < 0.05$, indicating a significant difference between the two distributions. Furthermore, these results are consistent with the findings of \cite{Popov..2002}, which analyzed PSR B1933+16 using the NASA Deep Space Network's 70-meter radio telescope at observation frequencies of 1634 MHz and 1650 MHz, reporting a characteristic micropulse width of 150 $\mu$s.

The width of micropulses is calculated using the following equation:

\begin{equation}
\begin{aligned}
\hspace{5 mm}
   \delta = \sqrt{\frac{ {\textstyle \sum_{i} } M_{i}(N_{i}-N_0)^{2}}{ {\textstyle \sum_{i}M_{i}} } } .
   \end{aligned}
	\label{eq:1}
\end{equation}

where the sum was performed over each micropulse in question. M$_i$ is the the amplitude of N$_i$, N$_i$ is the phase bin in ms. N$_0$ is the mean phase bin which can be obtained from

\vspace{-10pt}
\begin{figure*}[ht]  
\begin{equation}  
\begin{aligned}  
\hspace{5 mm}  
N_{0} &= \frac{ {\textstyle \sum_{i}}(M_{i} \cdot N_{i}) }{ {\textstyle \sum_{i}}M_{i} } \\  
\end{aligned}  
\label{eq:2}  
\end{equation}  
\end{figure*}

\begin{figure}[ht!]
 \centering
 \includegraphics[width=0.6\textwidth,height=0.35\textheight]{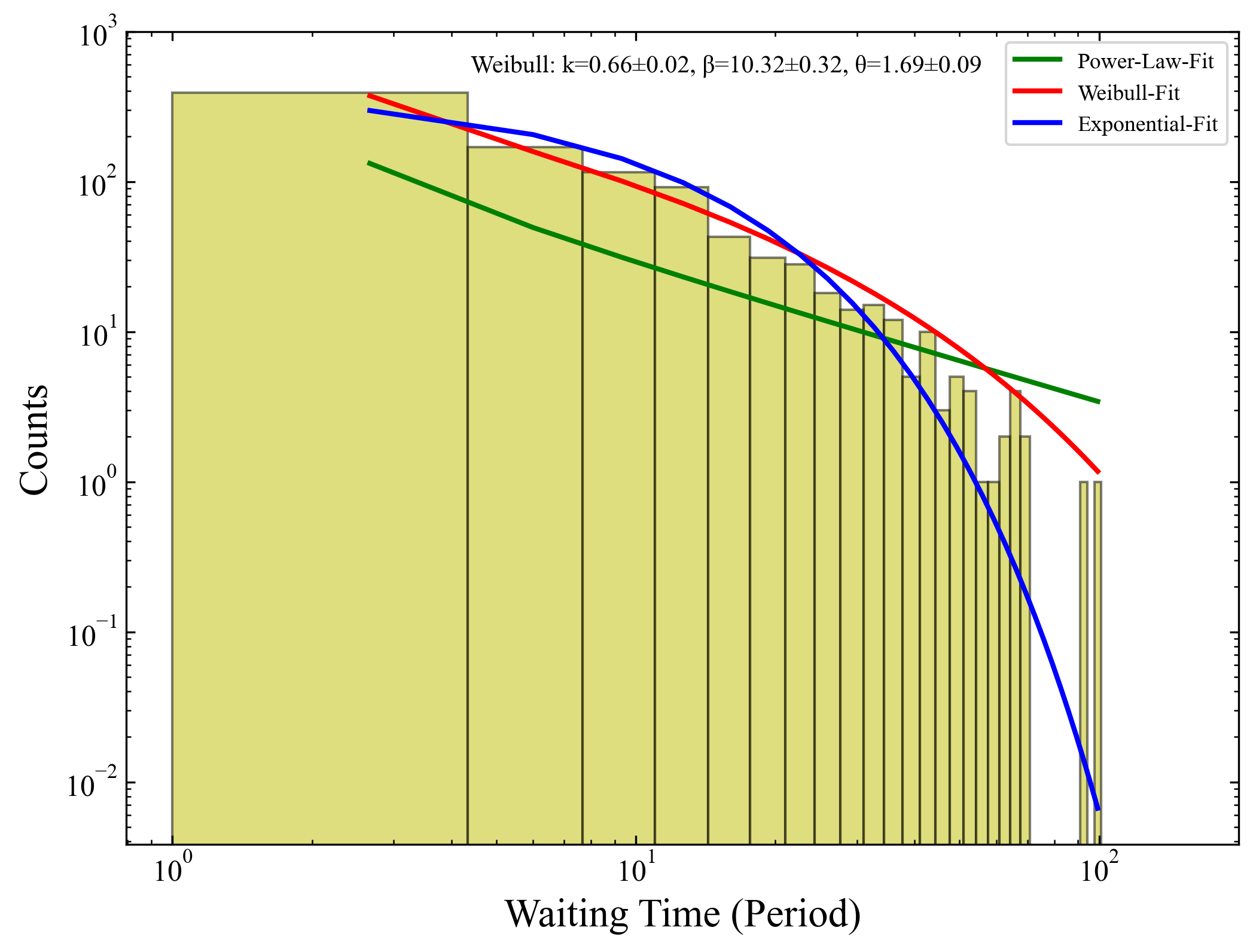}
\caption{
Fits to the waiting time distribution of 969 MPs from PSR J1935+1616 considering a power-law distribution (green curve), Weibull distribution (red curve), and exponential distribution (blue curve), with the main parameters of the Weibull distribution labeled above the figure.}
\label{waiting time}
\end{figure}

\begin{figure}[ht!]
 \centering
\includegraphics[width=0.85\textwidth,height=0.25\textheight]{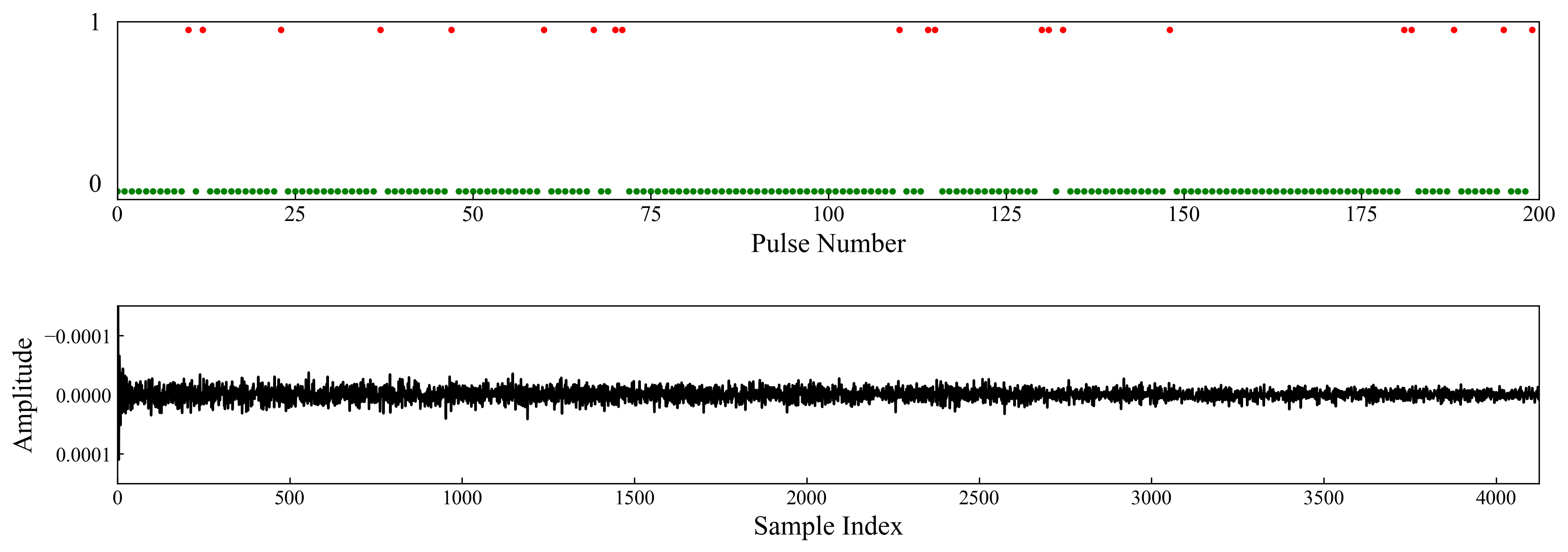}
\caption{The 9,998 single pulses of PSR J1935+1616 are represented as a binary time series resembling 1 and 0. The upper panel shows 200 of these pulses, with red dots indicating MPs (marked as 1) and green dots indicating normal pulses (marked as 0). The lower panel displays the result of the Fourier transform applied to the binary time series of all 9,998 single pulses.}
\label{period}
\end{figure}

\vspace{10mm}
\subsubsection{Time sequence analysis and 
energy distribution of MP, NP, TP}
\label{title_3.1.3}

To investigate the correlation between MPs, we calculated the time intervals (waiting time, $\Delta t$) between 9,998 single pulses. The waiting time statistic is widely used in the analysis of giant micropulse intervals \citep{2020ApJ...899..118C}, neighboring arrival times of fast radio bursts (FRBs) \citep{2022RAA....22l4002Z}, and solar flares \citep{2000SPD....31.0256W}, providing important information about the mechanisms of individual events.
The histogram of waiting time shown in Figure~\ref{waiting time} reveals a predominance of short waiting time ($<5$ periods), while the longest intervals between single pulses without micropulses can extend up to 101 periods. This indicates that the production of single pulse with microstructure (MP) is clustered, which leads us to use a Weibull distribution for the MP waiting time:

\begin{equation}
\begin{aligned}
\hspace{5 mm}
f(\Delta t) = \frac{k}{\beta} \left(\frac{\Delta t - \theta}{\beta}\right)^{k-1} e^{-\left(\frac{\Delta t - \theta}{\beta}\right)^k}, \quad \Delta t \geq \theta
\end{aligned}
\label{eq:3}
\end{equation}

where $\beta$ is the reciprocal of the mean occurrence rate. Using maximum likelihood estimation, the best-fitting coefficients are $k = 0.66 \pm 0.02$, $\beta = 10.32 \pm 0.32$, $\theta = 1.69 \pm 0.09$. $k < 1$ indicates that the probability of MP occurrence decreases over time, meaning that MP occurrences are clustered. $\theta > 1$ means that the waiting time for a MP is at least 1 period before $\Delta t = 1$. The exponent value in the power-law distribution determines the slope of the distribution. A smaller exponent suggests that the waiting time is generally short, but a few events can be very long, consistent with the characteristics of a long-tailed distribution. A simple Poisson process, where the probability of MP occurrence does not change over time, would produce an exponential distribution. It is evident that among the three fitted curves, the Weibull fit performs the best.

To study the periodicity of MP in the entire sample time series, we labeled MP as "1" and normal pulses as "0," converting all pulses into a binary-like pulsar time series. The Fourier transform of the entire pulse time series (see the bottom subplot of Figure~\ref{period}) indicates that the radiation process of MP is random. During the Fourier analysis of the single pulse sample data, we detected the presence of red noise. The spectral characteristic of red noise shows higher power at lower frequencies, which is typically associated with long timescale signal or noise features. To ensure that red noise does not affect our analysis results, we applied high-pass filtering to remove low-frequency noise and reduce its impact. In addition, we validated the robustness of the analysis results under different noise conditions. The experiments demonstrated that the presence of red noise has no substantial impact on the main conclusions.

\begin{figure*}
 \centering
\includegraphics[width=0.7\textwidth,height=0.55\textheight]{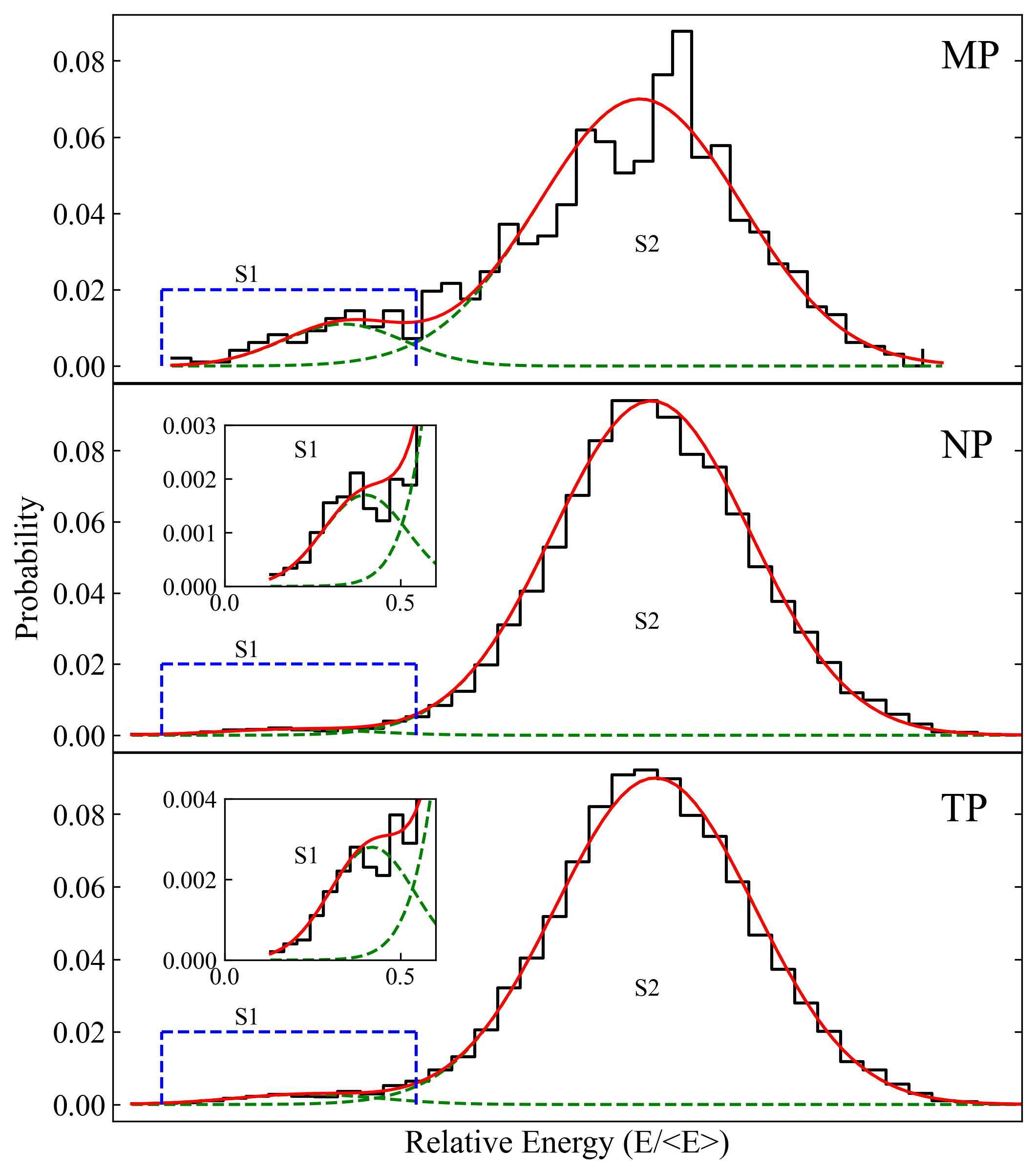}
    \caption{Histograms of the energy distributions for MP, normal pulses (NP), and total pulses (TP) of PSR J1933+1616. The black histograms represent the pulse energy distributions, the green dashed lines represent single Gaussian fit curves, and the red curves are the envelope of two Gaussian fit curves. S1 and S2 represent the two components of pulse energy. Since the S1 component is less prominent in NP and TP compared to MP, an enlarged view of this section is provided above the corresponding parts. The specific parameters of their Gaussian fit curves can be found in Table \ref{table:Guass}.}
    \label{mp_E}
\end{figure*}

We also analyzed the Relative Energy (E/<E>) distribution of MP, normal pulses (NP), and total pulses (TP), using the sum of flux values exceeding 3rms of the off-pulse as the single-pulse energy. The histograms for different pulses are shown in Figure \ref{mp_E}. Gaussian fitting was applied to each energy distribution, revealing a double structure, which we refer to as S1 and S2. Overall, MP, NP, and TP are  predominantly distributed in the S2 region, with a smaller portion in the S1 region. However, the proportion of S1 in MP is significantly higher than that in NP and TP. 

\subsubsection{\textbf{Polarization profile of MP, NP, TP }}\label{title_3.1.4}

\begin{figure*}
 \centering
\includegraphics[width=0.8\textwidth,height=0.6\textheight]{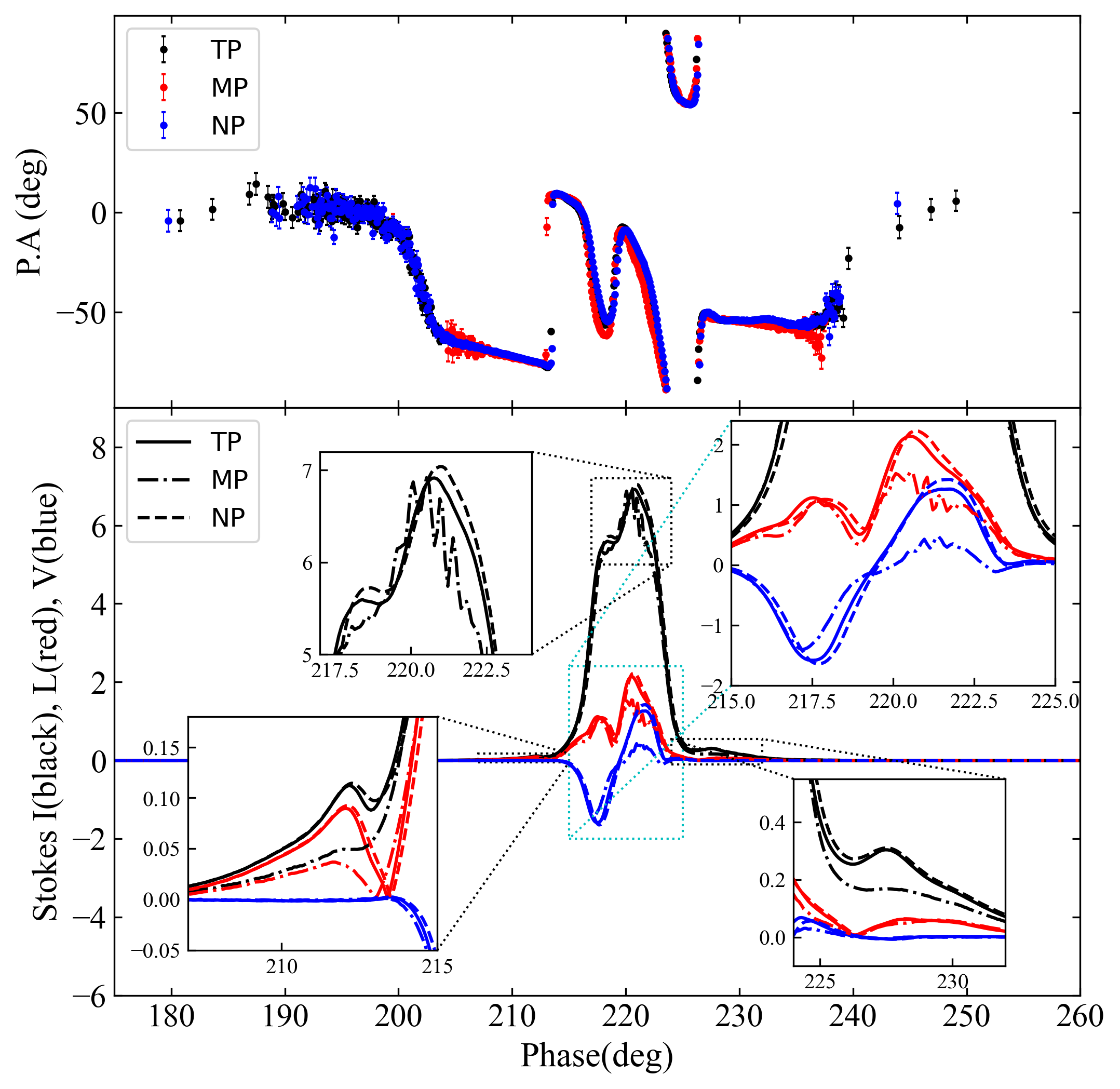}
    \caption{The upper plot illustrates the polarization position angles (PPA) of MP, NP, and TP. The lower plot shows the integrated profiles of polarization intensity, linear polarization, and circular polarization for MP, NP, and TP. Additionally, four inset plots provide zoomed-in views for a more detailed examination of specific areas.}
    \label{polaration.1}
\end{figure*}

Figure \ref{polaration.1} shows the comparison of average profiles between MP (dash-dotted line), normal pulses (NP, dashed line), and tatol pulses (TP, solid line). The intensity, linear polarization, and circular polarization profiles of TP are similar in structure to the profiles shown in Figure 2 of \cite{Mitra..2016MNRAS}. In the central component, the intensity of MP, NP, and TP is nearly identical. However, in the leading and trailing components, the intensity of MP is lower than that of NP and TP. 
Additionally, phase shifts are observed in the intensity profiles, compared to the peak phase of NP, the peak phase of MP tends to shift to the left.
In the leading component, the linear polarization of MP is lower than that of NP and TP, while the circular polarization is generally consistent across all three profiles. Between phases $215^\circ$ and $220^\circ$, the linear polarization of MP nearly overlaps with that of NP and TP, while the circular polarization is slightly higher. However, between phases $220^\circ$ and $225^\circ$, both the linear and circular polarization of MP are significantly lower than those of NP and TP. In the trailing component, the linear and circular polarization of MP shows small difference from NP and TP.
Furthermore, we observed (as shown by the "dot-dashed line" in Figure \ref{polaration.1}) that single pulses with microstructure exhibit modulation patterns similar to microstructures in the C2 component. These modulations are clearly visible in both the linear and circular polarization of the folded profile. The discovery is interesting, as it may suggest the existence of phase-locked microstructures, a phenomenon that has not yet been observed in pulsars. Further examination revealed that this phenomenon is not caused by a particularly bright single pulse with significant microstructure dominating the folded profile. Instead, it results from regional variations within the micropulse structure. However, the specific mechanism leading to this phenomenon remains unclear and requires further investigation.

The intensity of circular polarization in micropulses is lower than that of linear polarization, consistent with the findings of \cite{Mitra..2015ApJ} and \cite{singh..2023}.
The study by \cite{Rankin..1983ApJ} classifies pulsars based on their profile morphology and polarization characteristics, identifying two main types of pulsar emission: quasi-axial or "core" emission and the more common conal emission, with the triple profile being the most typical example. \cite{Xu..2000ApJ} and colleagues reproduced features like circular polarization reversal near the pulse center, S-shaped swings in linear polarization position angle, and strong linear polarization in conal components using the ICS model. This suggests the presence of a core component in MP, NP, and TP near phase $220^\circ$. Additionally, depolarization observed at the edges of linear polarization might relate to the conal structure, with the polarization position angle (PPA) of MP, NP, and TP generally remaining consistent during the pulsar's rotation.

\begin{figure*}
 \centering
\includegraphics[width=0.6 \textwidth,height=0.35\textheight]{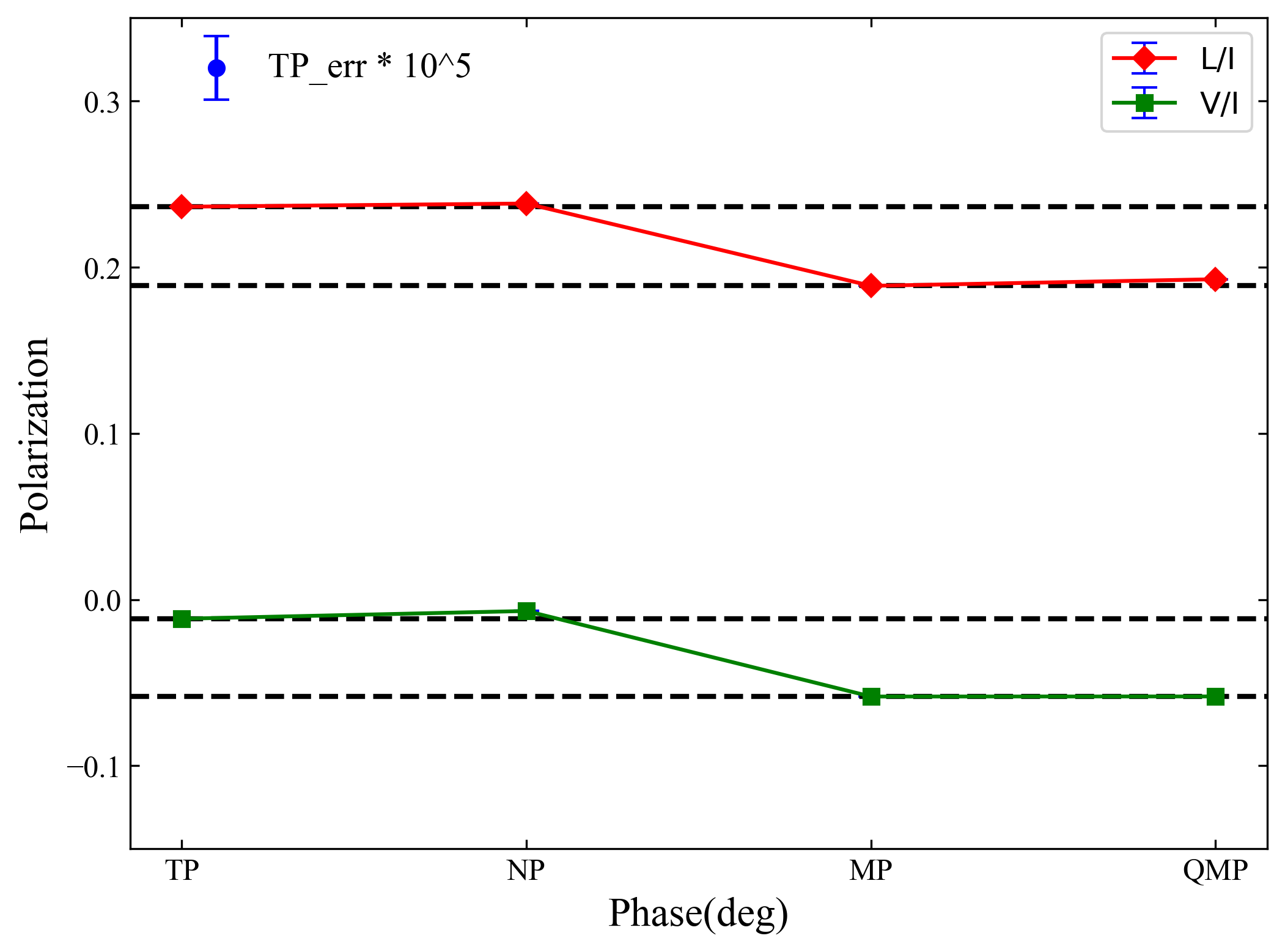}
    \caption{The polarization fractions of different types of pulses in PSR J1935+1616 where $L/I$ denotes linear polarization, and $V/I$ denotes circular polarization. TP stands for total pulses, NP stands for normal pulses, MP stands for \textbf{single pulse with} micropulses, and QMP stands for \textbf{single pulse with} quasi-periodic micropulses. Since the errors in these polarization fractions are very small, we have magnified the error in the linear polarization fraction of TP by a factor of $10^5$ and displayed it in the upper left corner of the figure. The specific error parameters for other types of pulses are detailed in Table \ref{table:P-err}.}
    \label{polaration.2}
\end{figure*}

\begin{table}[H]
\centering
\caption{Average polarization properties of the 9998 pulses from PSR J1936+1616 used in our study. \( \frac{L}{I} \) and \( \frac{V}{I} \) are the fractions of linearly and circularly polarised flux, and \( \delta\left(\frac{L}{I}\right) \times 10^{-7} \) and \( \delta\left(\frac{V}{I}\right) \times 10^{-7} \) are their uncertainties, respectively.}

\resizebox{0.7\textwidth}{!}{%
\begin{tabular}{cccccc}
\toprule
\diagbox{Name}{Parameters} & \( \frac{L}{I} \) & \( \frac{V}{I} \) & \( \delta\left(\frac{L}{I}\right) \times 10^{-7} \) & \( \delta\left(\frac{V}{I}\right) \times 10^{-7} \) \\
\midrule
TP  & 0.2364 & -0.0115 & 1.9100 & 2.0020 \\
NP  & 0.2383 & -0.0068 & 1.9153 & 2.1339 \\
MP  & 0.1888 & -0.0583 & 3.9065 & 6.0910 \\
QMP & 0.1928 & -0.0583 & 5.3822 & 8.1843 \\
\bottomrule
\end{tabular}
}
\label{table:P-err}
\end{table}

Figure \ref{polaration.2} shows the fractions of linear and circular polarization for TP, NP, MP, and QMP (specific parameters are provided in Table \ref{table:P-err}). It is evident that the circular polarization fractions are lower than the linear polarization fractions in all cases. When accounting for measurement errors, the linear and circular polarization fractions of TP and NP are consistent with those of MP and QMP. Additionally, the linear and circular polarization fractions of MP are lower than those of NP, with the rate of change in circular polarization being generally consistent with that of linear polarization.
We calculated the average profiles of intensity, linear polarization, and circular polarization for TP, NP, MP, and QMP, with the linear and circular polarization fractions based on the on-pulse regions of their average profiles. The errors were estimated using first-order partial derivatives. The formulas for calculating the linear and circular polarization fractions, as well as their error propagation formulas, are provided here.

The fractions of the linearly and circularly polarized flux are given by:

\begin{equation}
    \frac{L}{I} = \frac{\sqrt{Q^2 + U^2}}{I}
    \label{eq:linear_pol}
\end{equation}

where:
\( I \) is the total intensity,
\( Q \) and \( U \) are the Stokes parameters for linear polarization,
\( V \) is the Stokes parameter for circular polarization.

The standard deviations of the Stokes parameters \( I \), \( Q \), \( U \), and \( V \) are denoted as \( \sigma_I \), \( \sigma_Q \), \( \sigma_U \), and \( \sigma_V \), respectively. These standard deviations are calculated from the values of \( I \), \( Q \), \( U \), and \( V \) in the off-pulse region of the average profile. These standard deviations reflect the noise levels of each Stokes parameter in the off-pulse region. By using first-order partial derivatives, the propagation errors of the linear polarization fraction (\( L/I \)) and circular polarization fraction (\( V/I \)) can be further derived (The derivation process can be found in Appendix A).

\begin{figure*}
 \centering
\includegraphics[width=0.7 \textwidth,height=0.4\textheight]{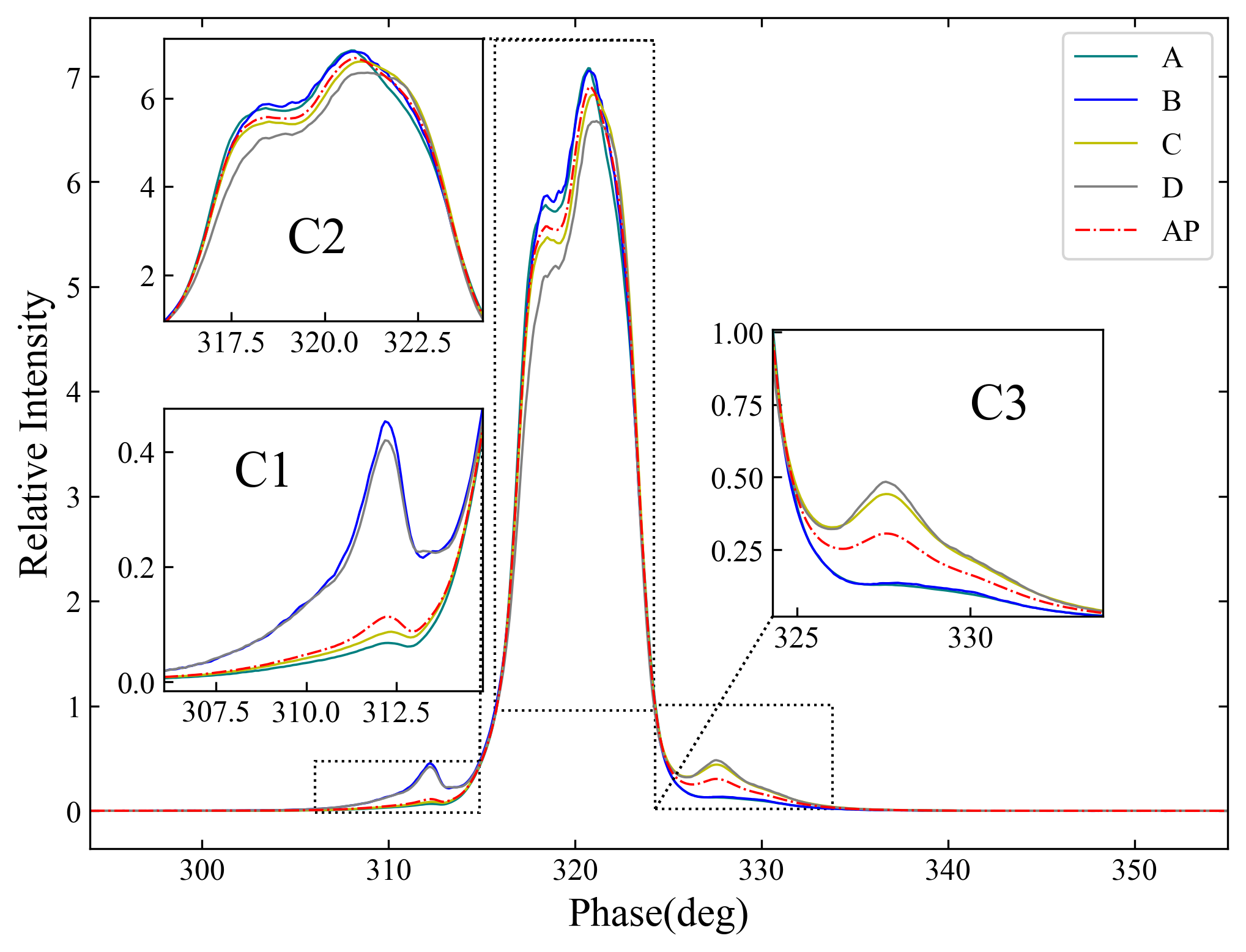}
    \caption{Categorization of modes A,B,C and D for PSRJ1935+1616 based on the integrated profile. The red curve (AP) represents the overall average profile of PSR J1935+1616, the cyan curve (A) represents the average profile for mode A, the blue curve (B) represents mode B, the yellow curve (C) represents mode C, and the gray curve (D) represents mode D.
   }
    \label{morphology}
\end{figure*}

\begin{figure*}
 \centering
\includegraphics[width=0.8 \textwidth,height=0.45\textheight]{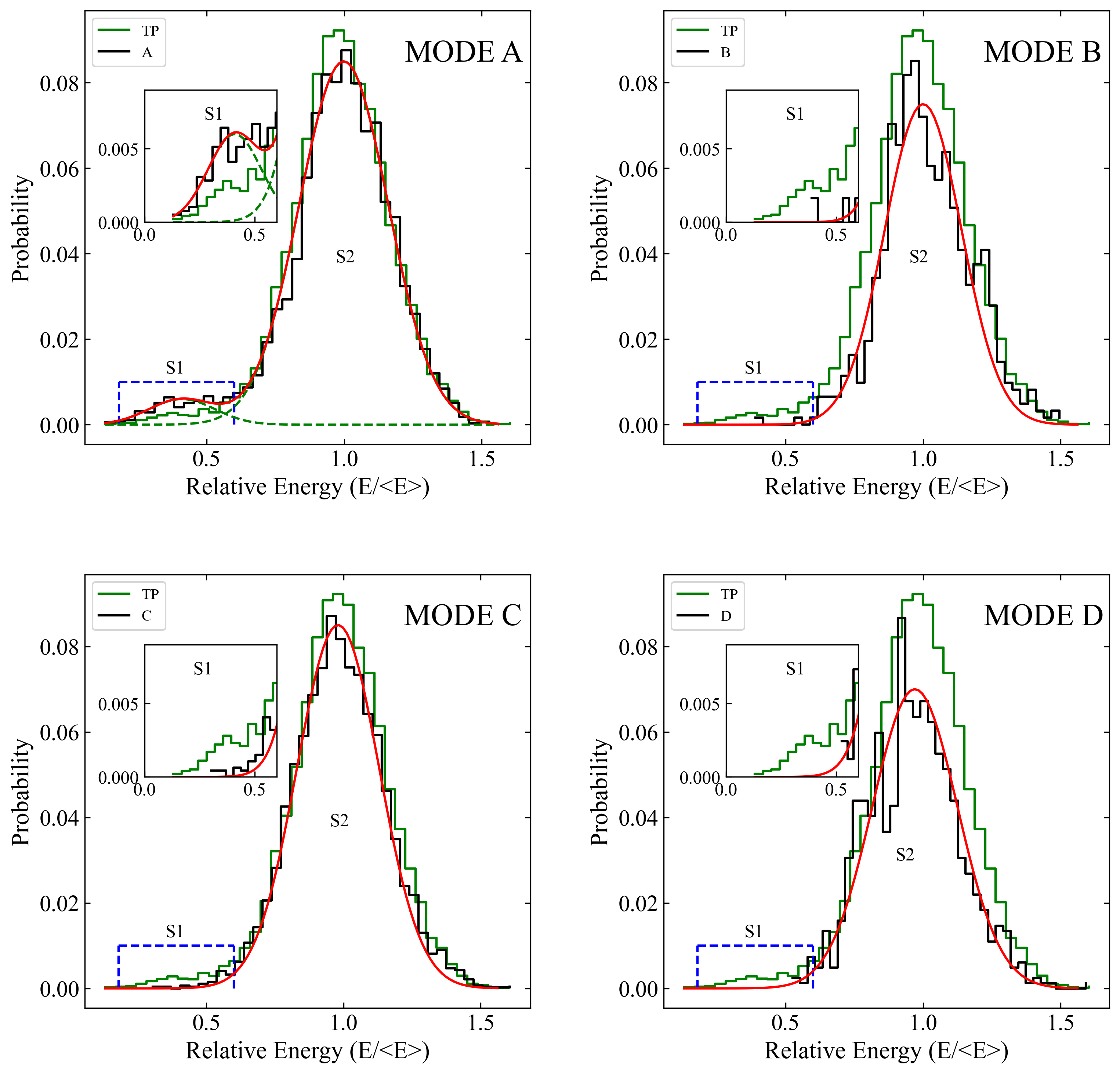}
    \caption{Based on the four single-pulse profile morphological modes (A, B, C, and D) of PSR J1935+1616, we analyzed the single-pulse energy distribution for each mode and performed Gaussian fits. The energy histograms are labeled with their respective mode names. The green histograms represent the energy distribution of total pulses (TP), while the black histograms show the energy distribution for each mode. The S1 region of each subplot provides a magnified view of the relative energy (E/<E>) in the range of 0.1-0.6. In the subplot for mode A, the green dashed lines represent the single Gaussian fit curves for the S1 and S2 energy regions, while the red curve is the envelope of these two Gaussian fit curves. In the subplots for modes B, C, and D, since the single-pulse energy distribution generally follows a single Gaussian distribution, we used a red curve to perform a single Gaussian fit for the respective energy distributions. The specific parameters of the Gaussian fit curves for the four modes are detailed in Table \ref{table:Guass}.
    }
    \label{morphology_E}
\end{figure*}

\vspace{5mm}
\subsection{Different Morphologies of Integrated Profiles and Energy Distributions of Single Pulses}\label{title_3.2}

The average pulse profile of PSR J1935+1616 consists of three components: the leading component (C1), the central component (C2), and the trailing component (C3) (see Figure \ref{morphology}). Based on this characteristic, the morphology of each single pulse can be divided into three components according to its phase position relative to the three components of the average profile. The intensity of 9,998 single pulses with different morphologies was then classified into four modes using a classification method based on the features of the average pulse profile components is employed:
\vspace{5mm}
\begin{itemize}
    \item \textbf{Mode A}: A single pulse is classified as mode A if the peak intensities of both its C1 and C3 components are lower than 10rms of the off-pulse.
    \item \textbf{Mode B}: A single pulse is classified as mode B if the peak intensity of its C1 component is higher than 10rms of the off-pulse, while the peak intensity of its C3 component is lower than 10rms of the off-pulse.
    \item \textbf{Mode C}: A single pulse is classified as mode C if the peak intensity of its C1 component is lower than 10rms of the off-pulse, while the peak intensity of its C3 component is higher than 10rms of the off-pulse.
    \item \textbf{Mode D}: A single pulse is classified as mode D if the peak intensities of both its C1 and C3 components are higher than 10rms of the off-pulse.
\end{itemize}

Figure \ref{morphology} shows the average pulse profile comparison of the four modes (the number of single pulses integrated for each mode is shown in Table \ref{table:Mode}). In mode A, the leading and trailing components are relatively weaker, while the central component is comparable to that in mode B and stronger than those in the other two modes. This mode lasts the longest in the entire sample's time series, reaching up to 9 single pulse periods. Mode B has a prominent leading component, while the trailing component almost disappears, similar to mode A, with a maximum duration of 6 single pulse periods. In mode C, the trailing component is more distinct than in modes A and B, and is comparable to that in mode D. The central component in mode C is weaker than in modes A and B, but stronger than in mode D, with the longest duration reaching 13 single pulse periods. In mode D, the leading and trailing components are the most prominent, while the central component is the weakest among the four modes, with a maximum duration of 5 single pulse periods. We performed Fourier analysis on the pulse sequences of different modes in the entire sample and found no periodicity in the changes of single-pulse morphology in any of the four modes. We used the sum of the flux values exceeding the 3rms of the off-pulse region as the energy of each single pulse. Then, by dividing the energy of each single pulse by the average energy of all pulses, we obtained the relative energy value of each single pulse. Figure \ref{morphology_E} shows the Relative Energy (E/<E>) distribution for each mode. The energy distribution of mode A is closer to the overall pulse energy distribution, exhibiting a double Gaussian distribution (S1 and S2 regions), while modes B, C, and D show a single Gaussian distribution.

\section{DISCUSSION}\label{title_4}
 
\subsection{Micropulse polarization characteristics}\label{title_4.2}

\begin{figure*}
 \centering
\includegraphics[width=0.7 \textwidth,height=0.4\textheight]{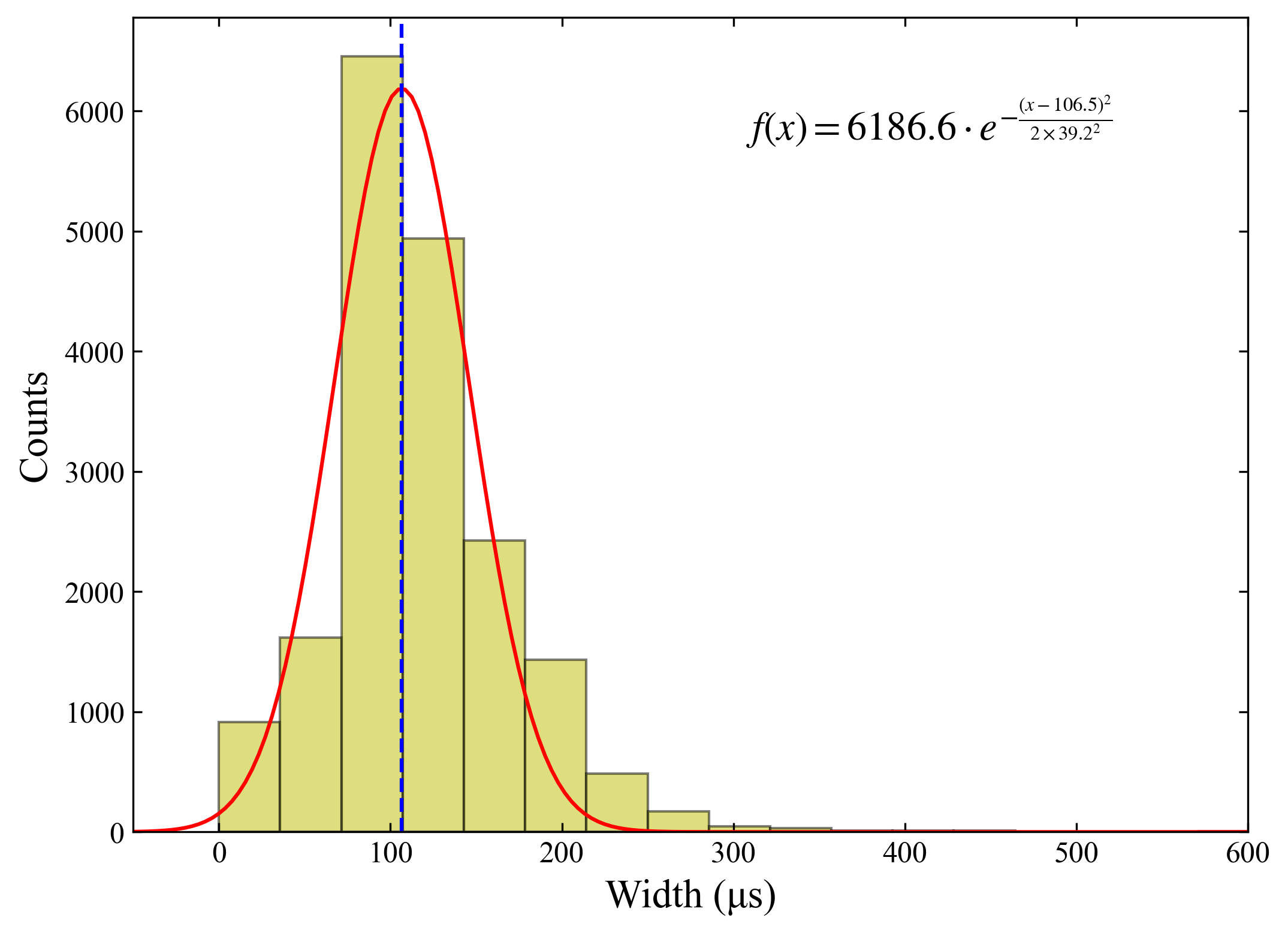}
    \caption{The histogram presents the width distribution of 969 single-pulse circularly polarized micropulses containing micropulses. The red solid line represents the best Gaussian fit, while the blue dashed line indicates the peak position of the Gaussian fit. The upper right corner displays the Gaussian fit equation along with the corresponding fitting parameters.
   }
    \label{V_Width}
\end{figure*}

According to \citep{Gil...1990,Tong_2022}, vacuum point-source curvature radiation typically causes a sign reversal in circular polarization and results in micropulses having a narrower width in circular polarization than in intensity. While
\cite{Mitra..2015ApJ} conducted a polarization analysis of 11 pulsars in the P-band and 32 pulsars in the L-band using the Arecibo telescope (with a time resolution of 59.5~$\mu$s) and found no clear evidence of point-source vacuum curvature radiation in microstructures. \cite{singh..2023} investigated the intrinsic micropulse polarization properties of PSR B0950+08 and B1642-03 using the Giant Metrewave Radio Telescope (GMRT) and found that the sign of circular polarization remained consistent within the micropulse window, further supporting this conclusion. Figure \ref{V_Width} presents the distribution of circular polarization micropulse widths for 969 single pulses containing micropulses, with a characteristic width of $106.52 \pm 46.14$  $\mu$s, which is consistent within the error range with the micropulse width in intensity (see Figure \ref{width} (a)). This result is consistent with the findings of \cite{Mitra..2015ApJ} and \cite{singh..2023}, suggesting that point-source vacuum curvature radiation is unlikely to be the dominant mechanism for micropulse emission.

\begin{table}[H]
\centering
\caption{Based on 1-hour observational data of PSR J1935+1616, we have compiled the distribution of the number of TP, MP, and QMP under different single-pulse morphological modes (A, B, C, and D).}
\resizebox{0.7\textwidth}{!}{%
\begin{tabular}{cccccc}
\toprule
\diagbox{Name}{Mode} &A &B &C &D &Total \\
\midrule
TP  &3893 &611 &4675 &819 &9998 \\
MP &562 &36 &351 &20 &969 \\
QMP &303 &19 &192 &6 &520 \\

\bottomrule
\end{tabular}
}
\label{table:Mode}
\end{table}

\subsection{Single pulse with micropulses and Single-Pulse Morphology}\label{title_4.3}

\begin{figure*}
 \centering
\includegraphics[width=0.6 \textwidth,height=0.35\textheight]{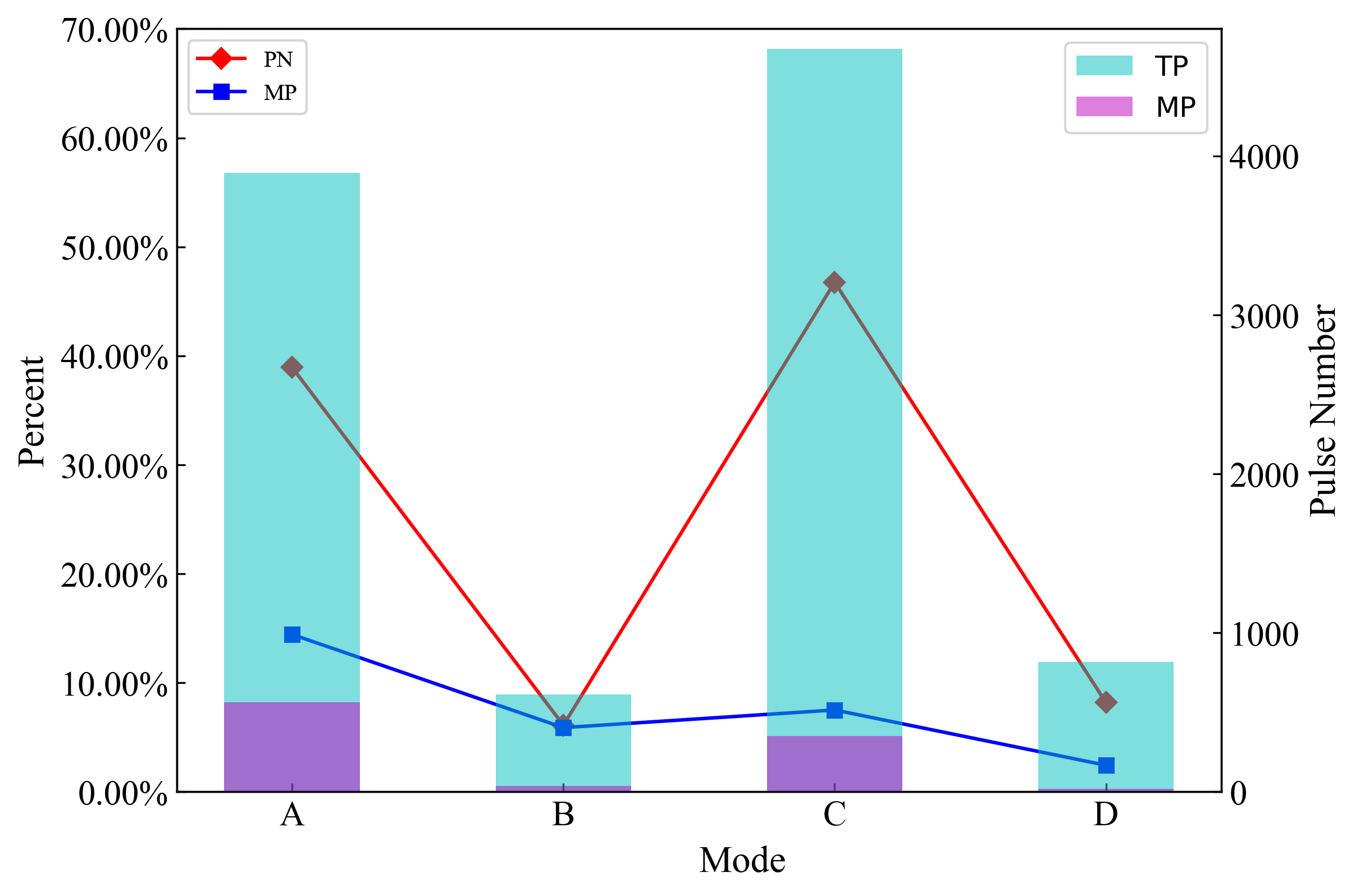}
    \caption{The statistical distribution and percentage trends of 969 MPs and 9,998 single pulses (TP) of PSR J1935+1616 across the four single-pulse morphologies (A, B, C, and D), with specific parameters detailed in Table \ref{table:Mode}.
}
    \label{statistic}
\end{figure*}

Table \ref{table:Mode} lists and gives the number of pulses, MPs, and QMPs in different single-pulse morphologies. Figure \ref{statistic} visually displays the histogram distribution and variation trends of the number of pulses, MPs, and QMPs in different single-pulse morphologies. From this Table \ref{table:Mode} and Figure \ref{statistic}, we can see that more than half of the MPs, QMPs are in mode A, while half of the single pulses are in mode C. Besides, when the pulse emission enters B and D modes, the MPs and QMPs are rarer, especially in D modes. This means that the MPs and QMPs are more likely emitted in mode A, in which the C2 component dominants the profile while C1 and C3 almost disappear. On the contrary, the numbers of MPs and QMPs are small in mode D, in which the outer components are obvious. 

\begin{figure*}
    \begin{minipage}{0.5\linewidth}
        \centering
        \includegraphics[width=\textwidth]{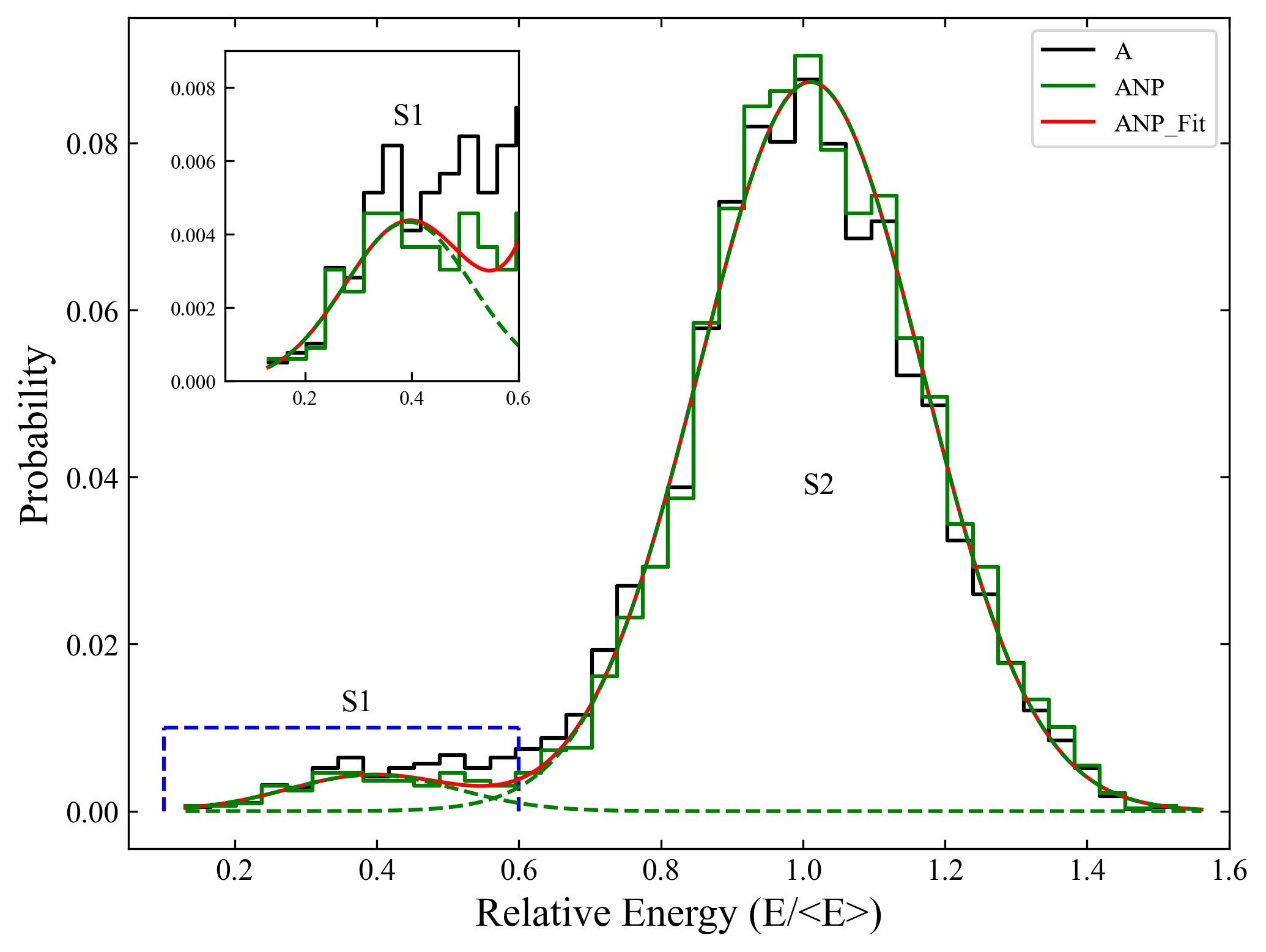}
        \centerline{(a)}
    \end{minipage}%
    \begin{minipage}{0.5\linewidth}
        \centering
        \includegraphics[width=\textwidth]{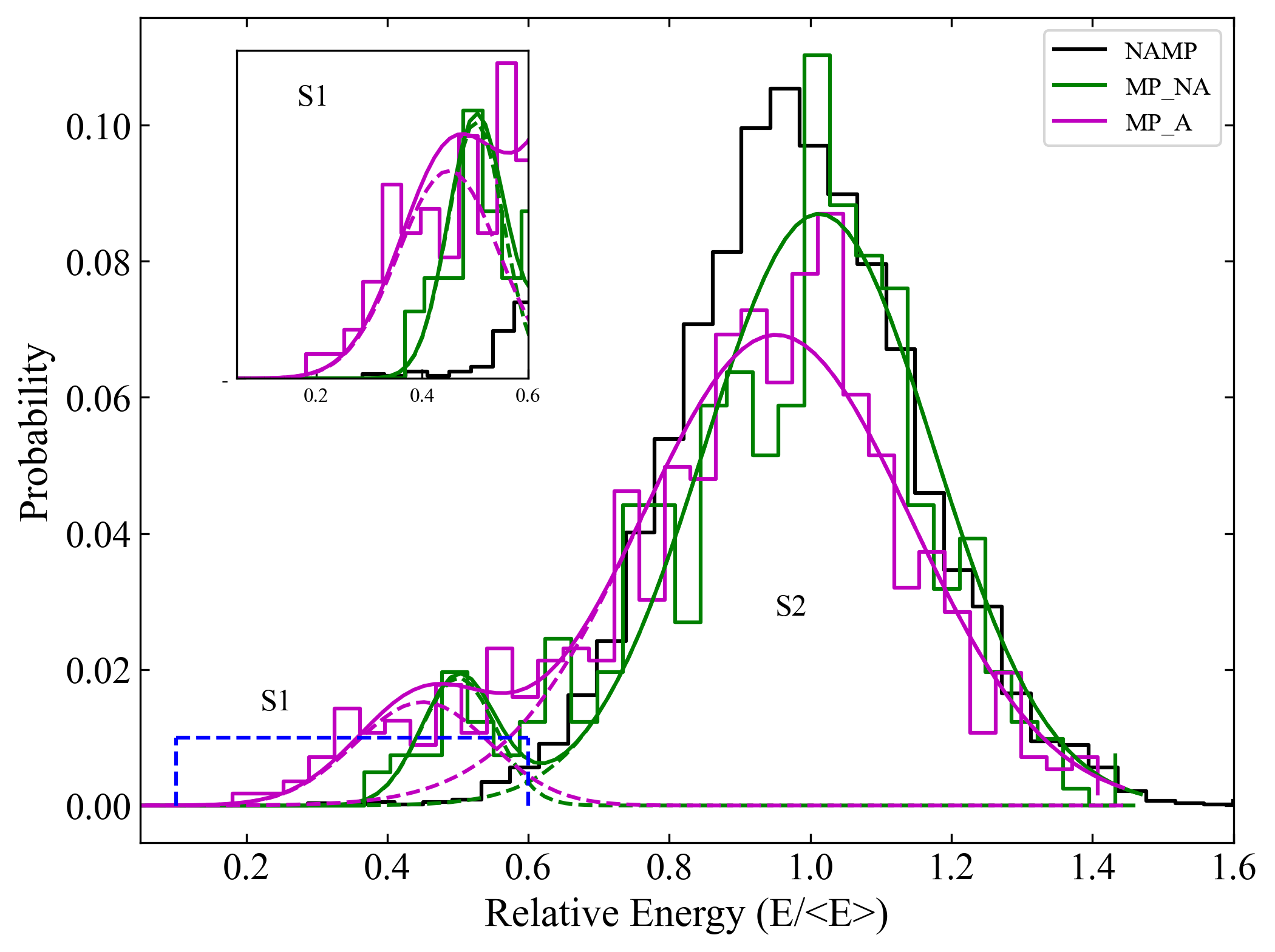}
        \centerline{(b)}
    \end{minipage}
    \caption{(a) Correspond to the energy histogram distribution of all pulses in Mode A (black histogram), which ANP represents the energy histogram distribution of normal pulses in Mode A (green histogram). The green dashed lines indicate the single Gaussian fit in the S1 and S2 regions for ANP, while the red curve represents the envelope of these two Gaussian fit (ANP\_Fit). (b) NAMP represents the energy histogram distribution of all pulses excluding Mode A and MPs (black histogram). MP\_NA represents the energy histogram distribution of MPs excluding those in Mode A (green histogram), with the green dashed lines indicating the single Gaussian fit curves in the S1 and S2 regions, and the green curve representing the envelope of these two green dashed Gaussian fit curves. MP\_A represents the energy histogram distribution of MPs in Mode A (magenta histogram), with the magenta dashed lines indicating the single Gaussian fit curves in the S1 and S2 regions, and the magenta curve representing the envelope of these two magenta dashed Gaussian fit curves. The specific parameters of the Gaussian fits are provided in Table \ref{table:Guass}.
    }
  \label{A_mode_E}
\end{figure*}

To validate the double Gaussian distribution of energy, Figure \ref{A_mode_E} presents two different histograms: (i) a comparison between MPs and single pulse without micropulses in mode A (Figure a); (ii) a comparison between mode A pulses and non-mode A pulses within MPs (Figure b). Additionally, Figure b shows the energy distribution of non-mode A and single pulse without micropulses. From Figure \ref{A_mode_E} (a), it can be observed that the energy of normal pulses in mode A (ANP) follows a Gaussian distribution in the relative energy range of 0.2 to 0.6, while a probability gap appears between the histograms of A and ANP in the relative energy range of 0.4 to 0.6, indicating that the energy of MPs in mode A is concentrated in this energy range. Figure \ref{A_mode_E} (b) shows that MP\_NA and MP\_A are significantly higher than NAMP when the relative energy is below 0.6. Combined with the comparison results of mode A in Figure a, this suggests that the double Gaussian distribution of energy is influenced by the radio emission of mode A and the presence of MPs. Moreover, Figure b indicates that low-energy MPs with relative energy below 0.35 are more likely to appear in mode A. Our conclusions are as follows: (i) The energy distribution of pulses and MPs in mode A both exhibit characteristics of a double Gaussian distribution; (ii) The overall pulse energy distribution presents a double Gaussian profile, which is attributed to the presence of mode A pulses and MPs.

\begin{table}[H]
\centering
\caption{The Gaussian fit curve parameters for the energy histogram distributions of MP,  NP, TP, Mode-A pulses, Mode-B pulses, Mode-C pulses, Mode-D pulses, ANP (normal pulses in Mode A), MP\_NA (MPs in non-Mode A), and MP\_A (MPs in Mode A). The energy distribution is divided into S1 and S2 regions, where the symbol ``dashed'' indicates the absence of that parameter. In the table, $A$ represents the amplitude of the Gaussian fit curve, $\mu$ represents the mean, and $\sigma$ represents the standard deviation.}
\resizebox{0.8\textwidth}{!}{%
\begin{tabular}{cccccccc}
\toprule
\diagbox{Name}{Region} & \multicolumn{3}{c}{S1} & & \multicolumn{3}{c}{S2} \\
\cline{2-4} \cline{6-8}
 & $A$ & \(\mu\) & $\sigma$ & & $A$ & \(\mu\) & $\sigma$ \\
\midrule
MP  & 0.0110 & 0.4800 & 0.1000 & & 0.0700 & 0.9700 & 0.1680 \\
NP & 0.0017 & 0.4000 & 0.1200 & & 0.0940 & 0.9890 & 0.1630 \\
TP & 0.0028 & 0.4200 & 0.1200 & & 0.0900 & 0.9950 & 0.1640 \\
Mode-A & 0.0060 & 0.4100 &0.1200 & & 0.0850 & 0.9980 & 0.1620 \\
Mode-B & $-$ & $-$ & $-$ & &0.0750 & 1.0000 & 0.1400 \\
Mode-C & $-$ & $-$ & $-$ & &0.0850 & 0.9800 & 0.1500 \\
Mode-D & $-$ & $-$ & $-$ & &0.0700 & 0.9700 & 0.1550 \\
ANP &0.0043 &0.3937 &0.1189 & &0.0873 &1.0108 & 0.1571 \\
MP\_NA &0.0188 &0.5013 &0.0528 & &0.0870 &1.0123 & 0.1622 \\
MP\_A &0.0152 &0.4503 &0.0949 & &0.0692 &0.9513 & 0.1922 \\
\bottomrule
\end{tabular}
}
\label{table:Guass}
\end{table}

\section{CONCLUSIONS}\label{title_5}

Based on FAST archive data, we obtained 9,998 single pulses of PSR J1935$+$1616 and discovered that this pulsar exhibits microstructure and changes in single-pulse morphology. We conducted a detailed analysis of the structural characteristics, phase clustering phenomena, periodicity analysis, polarization properties, and energy distribution of the single pulses with micropulses. Additionally, We summarized and analyzed the profile characteristics and energy distribution features of different single-pulse morphologies. This comprehensive study provides new insights into the behavior and underlying mechanisms of PSR J1935$+$1616, highlighting the complexity of its emission processes. In follow-up work we will continue to investigate the phase clustering phenomena of single pulses with micropulses, which will help to better understand the physical conditions and processes in the magnetosphere of pulsars exhibiting similar phenomena. The main conclusions of this study are as follows:

\begin{enumerate}
\item We identified 969 single pulses with micropulses, accounting for 9.69$\%$ of the total sample. These micropulses have a characteristic width of $127.63^{+70.74}_{-46.25}$ $\mu$s. Approximately 5.20$\%$ of these single pulses with micropulses exhibit quasi-periodicity, with a quasi-period of 231.77 $\pm$ 9.90 \(\mu\)s. The energy distribution of single pulses with micropulses is inconsistent with that of normal pulses, exhibiting a double Gaussian profile.
\item Among the 520 single pulses that exhibit quasi-periodic microstructure in intensity, 208 also show quasi-periodicity in circular polarization, with a characteristic quasi-periodicity of \(244.70^{+45.66}_{-21.05}\) $\mu$s. Additionally, among the 969 single pulses containing micropulses, the characteristic width of micropulses in circular polarization is \(106.52 \pm 46.14\) $\mu$s,  Which is consistent with their intensity width within the error range, indicating no polarization reversal within individual micropulses.

\item We found that PSR J1935$+$1616 exhibits four similar types of single-pulse profiles, each with distinct component intensity variations. As a result, we classified these profiles into four modes based on the pulse energy distribution characteristics.  Mode A's pulse energy follows a double Gaussian distribution, whereas Modes B, C, and D's pulse energy follow a single Gaussian distribution.
\item There is a relationship between single pulse with micropulses and different single-pulse morphologies. single pulses with micropulses are more likely to occur in mode A, while their probability in mode D is relatively low.
\end{enumerate}

\clearpage
\section{Acknowledgments}.
\begin{acknowledgments}
We gratefully acknowledge the financial support received from various funding sources for this work. This includes the National Natural Science Foundation of China (Grants No. 11988101, 12103013, U1731238, U2031117, 11565010, 11725313, 1227308, 12041303), the Foundation of Science and Technology of Guizhou Province (Grants No. (2021)023, (2016)4008, (2017)5726-37), the Foundation of Guizhou Provincial Education Department (Grants No. KY(2020)003, KY(2023)059), the National SKA Program of China (Grants No. 2022SKA0130100, 2022SKA0130104, 2020SKA0120200), the Youth Innovation Promotion Association CAS (id. 2021055), CAS Project for Young Scientists in Basic Research (grant YSBR-006), Foreign Talents Programme (Grant QN2023061004L, EG), the CAS Youth Interdisciplinary Team, the Liupanshui Science and Technology Development Project (NO.52020-2024-PT-01), and the Cultivation Project for FAST Scientific Payoff and Research Achievement of CAMS-CAS. Di Li is a New Cornerstone investigator. Their support has been instrumental in the successful completion of this work.
\end{acknowledgments}

%




\bibliographystyle{aasjournal}
\bibliography{references}{}

\appendix
\section*{Appendix A: Formula Derivation} 

\noindent
1. Error of Linear Polarization Fraction  \(\frac{L}{I}\)

For linear polarization fraction \(\frac{L}{I} = \frac{\sqrt{Q^2 + U^2}}{I}\), the partial derivatives with respect to \(Q\), \(U\), and \(I\) are:
\begin{align}
\frac{\partial (L/I)}{\partial Q} &= \frac{Q}{I \sqrt{Q^2 + U^2}}, \\
\frac{\partial (L/I)}{\partial U} &= \frac{U}{I \sqrt{Q^2 + U^2}}, \\
\frac{\partial (L/I)}{\partial I} &= -\frac{\sqrt{Q^2 + U^2}}{I^2}.
\end{align}

Thus, the error of the linear polarization fraction \(\sigma_{L/I}\) is:
\begin{equation}
\sigma_{L/I} = \sqrt{\left( \frac{\partial (L/I)}{\partial Q} \cdot \sigma_Q \right)^2 + \left( \frac{\partial (L/I)}{\partial U} \cdot \sigma_U \right)^2 + \left( \frac{\partial (L/I)}{\partial I} \cdot \sigma_I \right)^2}.
\end{equation}

Substituting the partial derivatives, we get:
\begin{equation}
\sigma_{L/I} = \sqrt{\left( \frac{Q \cdot \sigma_Q}{I \sqrt{Q^2 + U^2}} \right)^2 + \left( \frac{U \cdot \sigma_U}{I \sqrt{Q^2 + U^2}} \right)^2 + \left( \frac{\sqrt{Q^2 + U^2} \cdot \sigma_I}{I^2} \right)^2}.
\end{equation}

\vspace{1em} 

\noindent
2. Error of Circular Polarization Fraction  \(\frac{V}{I}\)

For circular polarization fraction \(\frac{V}{I}\), the partial derivatives with respect to \(V\) and \(I\) are:
\begin{align}
\frac{\partial (V/I)}{\partial V} &= \frac{1}{I}, \\
\frac{\partial (V/I)}{\partial I} &= -\frac{V}{I^2}.
\end{align}

Thus, the error of the circular polarization fraction \(\sigma_{V/I}\) is:
\begin{equation}
\sigma_{V/I} = \sqrt{\left( \frac{\partial (V/I)}{\partial V} \cdot \sigma_V \right)^2 + \left( \frac{\partial (V/I)}{\partial I} \cdot \sigma_I \right)^2}.
\end{equation}

Substituting the partial derivatives, we get:
\begin{equation}
\sigma_{V/I} = \sqrt{\left( \frac{\sigma_V}{I} \right)^2 + \left( \frac{V \cdot \sigma_I}{I^2} \right)^2}.
\end{equation}



\end{document}